\documentclass[traditabstract]{aa}  
\usepackage{graphicx}
\usepackage{epstopdf}
\usepackage{subfigure}
\usepackage{enumitem}
\usepackage{txfonts}
\usepackage{natbib}
\def\ltsima{$\; \buildrel < \over \sim \;$}
\def\simlt{\lower.5ex\hbox{\ltsima}}
\def\gtsima{$\; \buildrel > \over \sim \;$}
\def\simgt{\lower.5ex\hbox{\gtsima}}

\def\ergs{{erg s$^{-1}$}}
\def\cm2{{cm$^{-2}$}}

\def\f14{{10$^{-14}$}}
\def\f13{{10$^{-13}$}}
\def\f12{{10$^{-12}$}}
\def\f11{{10$^{-11}$}}

\def\uno{{J1326$-$0005}}
\def\due{{J1549$+$1245}}
\def\tre{{J1201$+$1206}}
\def\quattro{{J0745$+$4734}}
\def\cinque{{J0900$+$4215}}

\def\lbol{{L$\rm _{Bol}$}}

\def\kms{{km s$^{-1}$}} 
\def\msun{{$\rm M_{\rm \odot}$}}

\begin{document}

  \title{The WISSH Quasars Project}

   \subtitle{I. Powerful ionised outflows in hyper-luminous quasars}

   \author{M. Bischetti \inst{1, 2}
   \and E. Piconcelli \inst{1}
   \and G. Vietri\inst{1, 3}
   \and A. Bongiorno\inst{1}
   \and F. Fiore \inst{1}
   \and E. Sani \inst{4}
   \and A. Marconi \inst{5, 6}
   \and F. Duras \inst{7, 1}
   \and L. Zappacosta \inst{1}
   \and M. Brusa \inst{8, 9}
   \and A. Comastri \inst{9}
   \and G. Cresci \inst{6}
   \and C. Feruglio \inst{10, 1}
   \and E. Giallongo \inst{1}
   \and F. La Franca \inst{7}
   \and V. Mainieri \inst{11}
   \and F. Mannucci \inst{6}
   \and S. Martocchia \inst{1}
   \and F. Ricci \inst{7}
   \and R. Schneider \inst{1}
   \and V. Testa \inst{1}
   \and C. Vignali \inst{8, 9}
   }

   \titlerunning{Powerful ionised outflows in hyper-luminous quasars}
   \authorrunning{Bischetti et al.}
   
   \institute{INAF - Osservatorio Astronomico di Roma, Via Frascati 33, I--00078 Monte Porzio Catone (Roma), Italy
   \and Universit\`a degli Studi di Roma "Tor Vergata", Via Orazio Raimondo 18, I--00173 Roma, Italy
   \and Universit\`a degli Studi di Roma "La Sapienza", Piazzale Aldo Moro 5, I--00185 Roma, Italy
   \and European Southern Observatory (ESO), Alonso de Cordova 3107, Vitacura (Santiago), Chile
   \and Dipartimento di Fisica e Astronomia, Universit\`a di Firenze, Via G. Sansone 1, I--50019, Sesto Fiorentino (Firenze), Italy
   \and INAF - Osservatorio Astrofisico di Arcetri, Largo E. Fermi 5, I--50125, Firenze, Italy
   \and Dipartimento di Matematica e Fisica, Universit\`a degli Studi Roma Tre, via della Vasca Navale 84, I--00146, Roma, Italy
   \and Dipartimento di Fisica e Astronomia, Universit\`a di Bologna, viale Berti Pichat 6/2, I--40127 Bologna, Italy
   \and INAF - Osservatorio Astronomico di Bologna, via Ranzani 1, I--40127 Bologna, Italy
   \and INAF - Osservatorio Astronomico di Trieste, via G.B. Tiepolo 11, I--34143 Trieste, Italy 
   \and European Southern Observatory, Karl-Schwarzschild-str. 2, 85748 Garching bei M\"{u}nchen, Germany}
   \date{Accepted}

  \abstract
{Models and observations suggest that both power and effects of AGN feedback should be maximised in hyper-luminous ($\rm L_{Bol}>10^{47}$ \ergs) quasars, i.e. objects at the brightest end of the AGN luminosity function.
In this paper, we  present  the first results of a multi-wavelength observing  program, focusing  on a sample of WISE/SDSS selected hyper-luminous (WISSH) broad-line quasars at $z\approx 1.5-5$.
The WISSH quasars project has been  designed to reveal the most  energetic AGN-driven outflows, estimate their occurrence at the peak of quasar activity, and extend the study of correlations between outflows and nuclear properties up to poorly-investigated, extreme AGN luminosities, i.e. \lbol\ $\sim$ $10^{47}-10^{48}$ \ergs.
We present near-infrared, long-slit  LBT/LUCI1 spectroscopy of five WISSH quasars at $z\approx 2.3-3.5$, showing prominent [OIII] emission lines with broad (FWHM $\sim1200-2200$ \kms)  and skewed profiles.
 The luminosities of these broad [OIII] wings are  the highest measured so far, with $\rm L_{[OIII]}^{broad}\gtrsim5\times10^{44}$ \ergs,  and reveal the presence of powerful ionised outflows with associated mass outflow rates $\rm \dot{M}\gtrsim1700$ M$_\odot$ yr$^{-1}$ and kinetic powers $\rm \dot{E}_{\rm kin}\gtrsim10^{45}$ \ergs. Although these estimates are affected by large uncertainties, due to the use of [OIII] as tracer of ionized  outflows and the very basic outflow model we assume, these results suggest  that  the AGN is highly efficient in pushing outwards large amounts of ionised gas in our hyper-luminous targets. Furthermore, the mechanical outflow luminosities measured for WISSH quasars correspond to higher fractions ($\sim1-3$ \%) of \lbol\ than those derived for AGN with lower \lbol.
Our targets host very massive ($\rm M_{BH}\gtrsim2\times10^9$ $\rm M_\odot$) black holes which are still accreting at a high rate  (i.e. a factor of $\sim 0.4-3$ of the Eddington limit). These findings clearly demonstrate that WISSH quasars offer the opportunity of probing the extreme end of both luminosity and SMBH mass functions  and revealing powerful ionised outflows able to affect the evolution of their host galaxies.
}

 
 \maketitle
 %

 \section{Introduction}
  \vspace{-0.14cm}
  
 The energy output produced as a result of accretion on supermassive black holes (SMBHs) can be larger than the  gravitational binding energy in their host galaxies, and it is therefore expected to have a substantial role in their evolution \citep{SilkRees1998,WyitheLoeb2003,KingPounds2015}.
 Models of galaxy  formation indeed invoke Active Galactic Nuclei (AGN)-driven feedback to suppress the excessive growth of massive galaxies and drive their evolution from gas-rich starburst galaxies to red and dead ellipticals \citep{HopkinsHernquistCoxEtAl2008,SomervilleHopkinsCoxEtAl2008}.
 There is a general consensus in regarding AGN-driven outflows as the most efficient feedback mechanism to deplete and/or heat the interstellar medium (ISM) and, possibly, quench star formation \citep{MenciFiorePuccettiEtAl2008,ZubovasKing2012,CostaSijackiHaehnelt2014}.  
 
 Recently, a rapidly flourishing literature on AGN winds has been growing, with the discovery of multiple observational examples in all (i.e. molecular and atomic, neutral and ionised) gas phases \citep[e.g.][]{FeruglioFioreCarnianiEtAl2015,MaiolinoGalleraniNeriEtAl2012,Fabian2012,HarrisonAlexanderSwinbankEtAl2012,SpoonFarrahLebouteillerEtAl2013,RupkeVeilleux2015,BrusaBongiornoCresciEtAl2015,MorgantiVeilleuxOosterlooEtAl2016},  with kinetic powers clearly  indicating that these winds must originate from nuclear activity, consistently with the expectations of AGN feedback and galaxy co-evolution models \citep[see][and references therein]{FioreEtAl2016}.
 In particular,  the observation of emission lines, typically [OIII]$\lambda4959,5007$ \AA, with broad, skewed, blue/redshifted profiles has been the main tool to prove that ionised gas winds are a nearly ubiquitous feature in AGN \citep{ZhangDongWangEtAl2011,MullaneyAlexanderFineEtAl2013,ShenHo2014}.
 Thanks to  spatially-resolved spectroscopy, it has been possible to reveal galaxy-wide, kpc-scale [OIII] outflows in dozens of AGN
 \citep[e.g.][]{NesvadbaLehnertEisenhauerEtAl2006,NesvadbaLehnertDeBreuckEtAl2008,HarrisonAlexanderSwinbankEtAl2012,HarrisonAlexanderMullaneyEtAl2014,LiuZakamskaGreeneEtAl2013,CarnianiMarconiMaiolinoEtAl2015,PernaBrusaSalvatoEtAl2015} and  infer their mass outflow rate and  kinetic power, with measured $\rm \dot{M}$ and $\rm \dot{E}_{\rm kin}$ up to $\rm \sim 600-700$ $\rm M_{\odot}$ yr$^{-1}$ and $\sim0.01\times$ \lbol, respectively.
 Furthermore, in some cases,  the outflowing [OIII] emission  has been found to be spatially anti-correlated with narrow H$\alpha$ emission, which traces star formation in the host galaxy \citep{Cano-DiazMaiolinoMarconiEtAl2012,CresciMainieriBrusaEtAl2015,CarnianiMarconiMaiolinoEtAl2016}. These results are considered among the  most compelling pieces of evidence of negative feedback from AGN reported so far.
 

 Remarkably, both models and observations indicate that the AGN efficiency in driving energetic winds and the momentum fluxes of galaxy-scale outflows increase with AGN luminosity \citep[e.g.][]{MenciFiorePuccettiEtAl2008,FaucherGiguereQuataert2012,CiconeMaiolinoSturmEtAl2014,FeruglioFioreCarnianiEtAl2015,HopkinsTorreyFaucherGiguereEtAl2016}. 
 Understanding  the coupling between the nuclear energy output and the ISM in the host galaxy is thus particularly relevant in the context of  hyper-luminous AGN (i.e. quasars with $\rm L_{Bol}$ $>$ 10$^{47}$ \ergs), lying at the  bright end of the AGN luminosity function.
 A systematic,  multi-wavelength investigation of  nuclear, outflows, star-formation and ISM properties  of hyper-luminous quasars is therefore clearly needed  to understand the impact of their huge radiative output on the host galaxy evolution.
 More specifically,  the role of AGN luminosity in accelerating winds and providing an efficient feedback, and the occurrence of powerful outflows in hyper-luminous systems at the peak epoch of quasar and star-formation activity (i.e. at z $\sim2-3$), are key aspects that deserve a detailed study in order to be answered.
 To address these open issues, we have designed a multi-band (from millimetre wavelengths up to hard X-rays) investigation of a  large sample of WISE/SDSS selected hyper-luminous (WISSH) quasars.

 In this paper,  which  is the first of a series that will focus on WISSH quasars, we present observations and physical parameters of AGN-driven  ionised outflows  traced by broad [OIII] emission lines, which have been detected in the rest-frame optical spectra of five  quasars at $z\sim2.3-3.5$. 
 In Sect. \ref{sec:WISSH}, we describe the WISSH quasars sample and the main goals of our on-going project. Details about LBT observations and data reduction are presented in Sect. \ref{sec:obs}. 
 In Section \ref{sec:analysis}, we measure the extinction  and   outline the  models used in our spectral analysis.
 In Section \ref{sec:results},  we present the results of the spectral fitting, i.e.  the best-fit models and emission lines parameters, and  the analysis of the "off-nuclear" spectra, performed to detect possible presence of extended [OIII] emitting gas. 
 H$\beta$-based SMBH masses and Eddington ratios are presented in Sect. \ref{sec:smbh}.
 In Sect. \ref{sec:discuss}, we discuss the properties of the broad [OIII] emission derived from our observations  in the context of luminous quasars, and we compare the mass rate and kinetic power of the outflows of WISSH quasars  with those found for lower-luminosity AGN samples.
 We summarise our findings in Sect. \ref{sec:conclusions}. 
 
 Throughout this paper, we assume  $\rm H_0$ = 70 km s$^{-1}$ Mpc$^{-1}$, $\rm \Omega_\Lambda$ = 0.73 and  $\rm \Omega_M$= 0.27.

\section{The WISSH quasars project}\label{sec:WISSH}

\begin{table*}[t]
	\centering
	\caption{Properties of the five analysed WISSH quasars. Columns give the following information: (1) SDSS ID,  (2--3) celestial coordinates, (4) redshift from the SDSS DR10 catalog, (5--9) photometric data from the SDSS DR10 catalog, (10--12) photometric data from the 2MASS catalog.}
	\makebox[\columnwidth]{
		\begin{tabular}{lccccccccccc}
			
			\hline
			SDSS & RA & Dec  & z$_{\rm SDSS}$ & \it{u} & \it{g} & \it{r} & \it{i} & \it{z} & J & H & K$_{\rm s}$ \\ 
			(1) & (2) & (3) & (4) & (5) & (6) &(7) & (8) & (9) & (10) & (11) & (12) \\
			\hline
			J0745$+$4734 & 07:45:21.78 & $+$47:34:36.1 &  3.214 &  19.56 & 16.63 & 16.35 & 16.29 & 16.19 & 15.08 & 14.61 & 13.95 \\
			J0900$+$4215 & 09:00:33.50 & $+$42:15:47.0 &  3.295 &  22.97 & 17.11 & 16.74 & 16.69 & 16.58 & 15.37 & 14.68 & 14.00 \\
			J1201$+$1206 & 12:01:47.90 & $+$12:06:30.3 &  3.484 &  20.77 & 18.31 & 17.41 & 17.31 & 17.18 & 15.87 & 15.25 & 14.61 \\
			J1326$-$0005 & 13:26:54.95 & $-$00:05:30.1 &  3.307 &  22.83 & 20.90 & 20.54 & 20.02 & 19.28 & 17.39 & 16.75 & 15.31 \\ 
			J1549$+$1245 & 15:49:38.71 & $+$12:45:09.1 &  2.386 &  20.23 & 18.67 & 17.84 & 17.38 & 16.90 & 15.86 & 14.56 & 13.52 \\

			\hline 
		\end{tabular} 
	}
	\label{tab:mag}
\end{table*}

\begin{table*}[]
	\centering
	\caption{Journal of the LBT/LUCI1 observations. Columns give the following information: (1) SDSS ID,  (2)  grating, (3) resolution, (4) exposure time (in units of s), (5) average seeing (in units of arcsec), (6) observation date, and (7) standard star.}
	\makebox[\columnwidth]{
		\small
		\begin{tabular}{lcccccc}
			
			\hline
			SDSS &  Grating & R & t$_{\rm exp}$   & Seeing & Obs date & Std. Star\\ 
			(1) & (2) & (3) & (4) & (5) & (6) & (7) \\
			\hline
			J0745$+$4734 & 150 K$_s$ & 4150 & 600  & 0.90 & Jan 10\textsuperscript{th} 2015 & HIP35787 \\
			J0900$+$4215 & 150 K$_s$ & 4150 & 800  & 1.0  & Jan 10\textsuperscript{th} 2015 & HIP35787 \\
			J1201$+$1206 & 150 K$_s$ & 4150 & 1500 & 0.72 & Apr 20\textsuperscript{th} 2014 & HIP64231 \\ 
			J1326$-$0005 & 150 K$_s$ & 4150 & 2500 & 0.79 &  Apr 20\textsuperscript{th} 2014 & HIP64231 \\ 
			J1549$+$1245 & 210 zJHK  & 7838 & 1500 & 0.66 &  Apr 20\textsuperscript{th} 2014 & HIP68868 \\ 
			
			\hline 
		\end{tabular} 
	}
	\label{tab:data}
\end{table*}

We are following up the properties of a quasars sample compiled by \cite{WeedmanSargsyanLebouteillerEtAl2012}, consisting of 100 Type 1 AGN at $z\gtrsim 1.5$ with a flux density S$_{22 \mu m}>3$ mJy, selected by cross-correlating  the WISE all-sky source catalogue \citep{WrightEisenhardtMainzerEtAl2010} with the SDSS DR7 catalogue. Our sample of WISSH quasars has been obtained by removing lensed objects and those with a contaminated WISE photometry from the original list of \cite{WeedmanSargsyanLebouteillerEtAl2012} and consists of 86 quasars at $z\sim1.8-4.6$.
As stressed by \cite{WeedmanSargsyanLebouteillerEtAl2012}, WISSH quasars represent the most luminous AGN known in the Universe, with  bolometric luminosities \lbol\ $>10^{47}$ erg s$^{-1}$. By selecting Type 1 SDSS quasars, we include sources affected by low extinction and for which it is possible to reliably measure the SMBH mass via broad emission lines width. The vast majority of the existing works focused on AGN-driven feedback has been concentrated on AGN samples with redshifts and luminosities smaller than  those of WISSH sources  \citep[e.g.][]{HarrisonAlexanderSwinbankEtAl2012,HarrisonAlexanderMullaneyEtAl2016,RupkeVeilleux2013,WylezalekZakamska2016}.
The WISSH survey is specifically designed to systematically investigate the outflows properties in \lbol\ $>10^{47}$ erg s$^{-1}$ quasars at cosmic noon, i.e. $\rm z \sim$2 -- 3, when SMBH accretion and star formation reached their maximum, and complement previous literature on observational evidence for AGN feedback.

The main goal of our WISSH project is probing the properties of nuclear and star formation activity in these  hyper-luminous quasars, as well as the effects of AGN feedback on their host galaxies, via spectroscopic, imaging and photometric observations.
An extensive, multi-band, observing program is indeed currently on-going, based on ALMA, SINFONI, X-Shooter, TNG, XMM-Newton  and  Chandra data (the project also benefits from publicly available  Herschel, WISE, 2MASS and SDSS data) in order to obtain a panchromatic, less-biased view of
these extreme sources. Such a large amount of data indeed enables us to properly build the spectral energy distribution (SED),  investigate the  different  ISM phases and constrain the  SMBH  mass and accretion rate of WISSH quasars. Furthermore,  WISSH quasars allow to improve the study of correlations between outflows and nuclear properties, by bringing the explorable range up to  \lbol\ $\sim$ $10^{47-48}$ erg s$^{-1}$.

\section{LBT observations and Data Reduction} \label{sec:obs}
In this paper, we present the first results of the WISSH quasars project.
We have been awarded observing time with the Near-Infrared Spectrograph and Imager  (LUCI1) at the 8.4 m Large Binocular Telescope (LBT), located on Mount Graham (Arizona), to perform moderate resolution (R $\sim$ 4000) near-IR spectroscopy  of 18 WISSH quasars, with the aim of revealing possible broad/blueshifted [OIII] components, indicative of ionised outflows. These objects belong to a subsample of WISSH quasars selected by removing sources with (i) $z>3.6$ and $1.6< z<2$ (i.e. for which LUCI1 does not cover the
[OIII] spectral region), (ii) no available 2MASS JHK magnitudes, and (iii) a redshift for which telluric lines contaminate the [OIII] emission
line or the LBT/LUCI1 bandwidth does not simultaneously cover both [OIII] and
H$\beta$ emission lines. Eight targets were observed during Cycle 2014, while the remaining ones belong to Cycle 2015. 

Here we focus on the five quasars (namely \quattro, \cinque, \tre, \uno\ and \due),  which exhibit prominent, broad [OIII] emission. The complete analysis and results for the whole LUCI1 WISSH quasars sample, consisting of 18 objects, will be reported in a forthcoming paper (Vietri et al.). Table \ref{tab:mag} lists coordinates, SDSS 10\textsuperscript{th} Data Release (DR10) redshift and optical photometric data \citep{AhnAlexandroffAllendeEtAl2014} and 2MASS near-IR photometric data \citep{StruskieCutriStieningEtAl2006} for the five WISSH quasars analysed here.

The LUCI1 observations of our targets were carried out in long-slit mode (1 arcsec slit at position angle PA = 0 deg from North) and with the N1.8 camera,  during the nights of April 20\textsuperscript{th} 2014 and January 10\textsuperscript{th} 2015.  We used the grating 150\_K$\rm_s$ ($\rm R=4140$) for the  $z>3$ quasars and the grating 210\_zJHK ($\rm R=7838$) in the H-band for \due\   ($z \sim 2.4$). The average seeing during the observations was $\sim0.7-0.8$ arcsec for Cycle 2014 and $\sim0.9-1.0$ arcsec for Cycle 2015 (see Table \ref{tab:data}). The spectra of telluric standards (A0V--A1V stellar types, see Table \ref{tab:data}) were  acquired  immediately before/after the quasar.  

Data were reduced using standard IRAF packages in combination with IDL tasks.
The two-dimensional (2D) wavelength calibration was carried out on each frame of the five targets using Argon and Neon lamp exposures for \quattro\ and \cinque, Argon and Xenon lamp exposures for \due\ and sky emission lines for \tre\ and \uno. 
The sky background subtraction was performed using the nodding technique.
From the 2D calibrated and sky lines free spectra, we extracted one-dimensional (1D) spectra for both target and telluric standard star.
As a final step, the flux calibration and telluric absorption correction were applied to the 1D spectra using the IDL based routine XTELLCOR$\_$GENERAL \citep{VaccaCushingRayner2003}.  This routine employs a high-resolution spectrum of Vega, convolved with the instrumental resolution, to fit  both telluric and stellar features in the standard star spectrum, in order to compute the  telluric absorption correction for the quasar spectrum.
The flux calibration, as a function of wavelength,  was performed  by  a response curve based on the telluric standard B and V magnitudes, including the reddening curve of \cite{RiekeLebofsky1985}.

\section{Analysis}\label{sec:analysis}

\subsection{Spectral analysis}
\label{sec:spectral}
\begin{figure*}[t]
	\centering
	\caption{LUCI1 spectra of the five analysed WISSH quasars. The red lines show the resulting best-fits for \textit{(a)} SDSS \quattro, \textit{(b)} \cinque, \textit{(c)} \tre, \textit{(d)} \uno\ and \textit{(e)} \due. Blue and green curves respectively refer to the NLR core and broad emission of both [OIII] and H$\beta$. BLR H$\beta$ emission is indicated in orange and FeII emission is indicated in magenta. The cyan curve in panel \textit{(c)} represents the additional components necessary to account for the \textit{plateau}-like emission feature between 4900-5000 \AA\ in \tre. Grey bands indicate the regions excluded from the fit due to presence of telluric features.}
	\makebox[1\textwidth]{
		\includegraphics[width=0.5\linewidth]{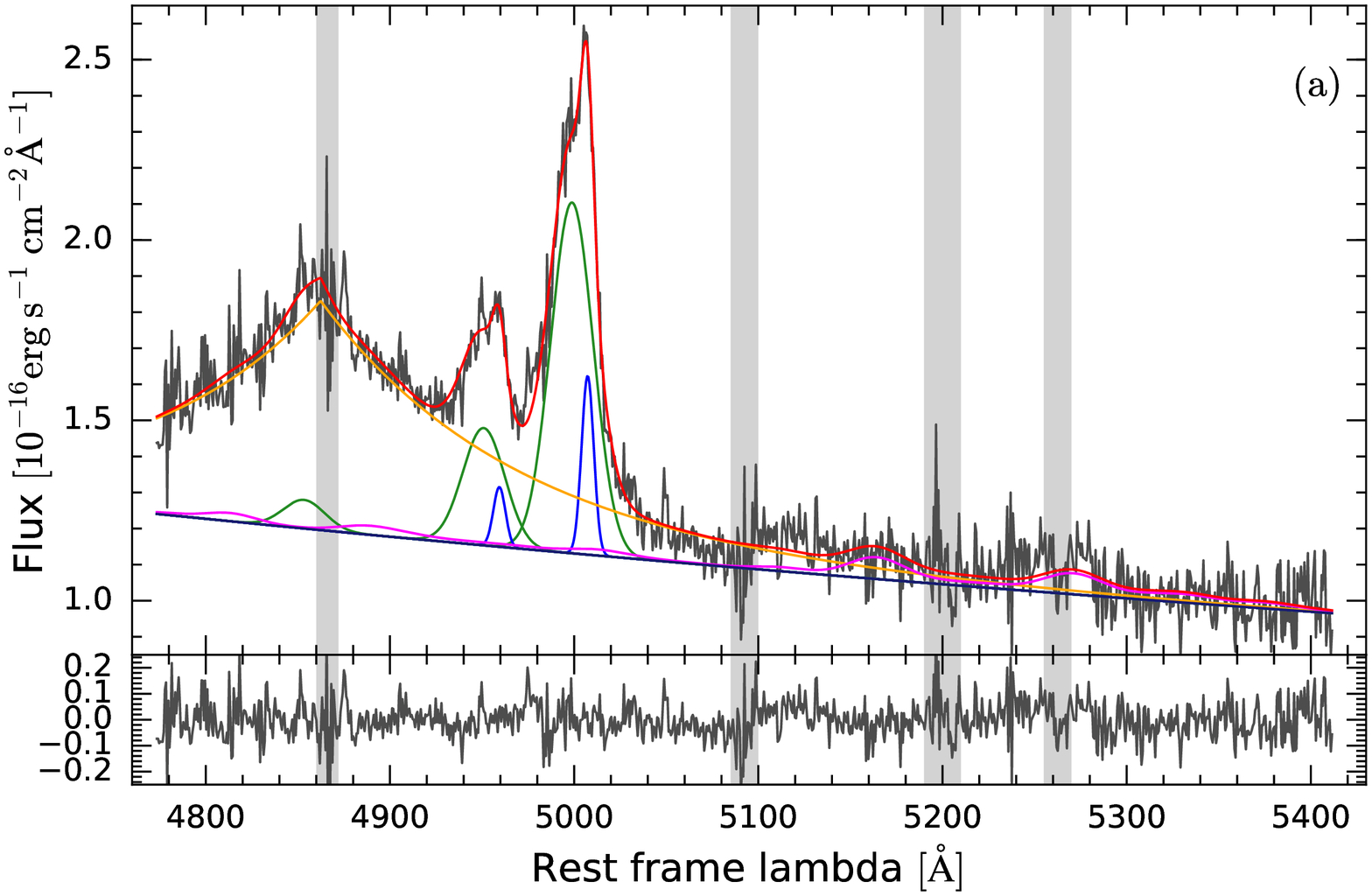}
		\includegraphics[width=0.5\linewidth]{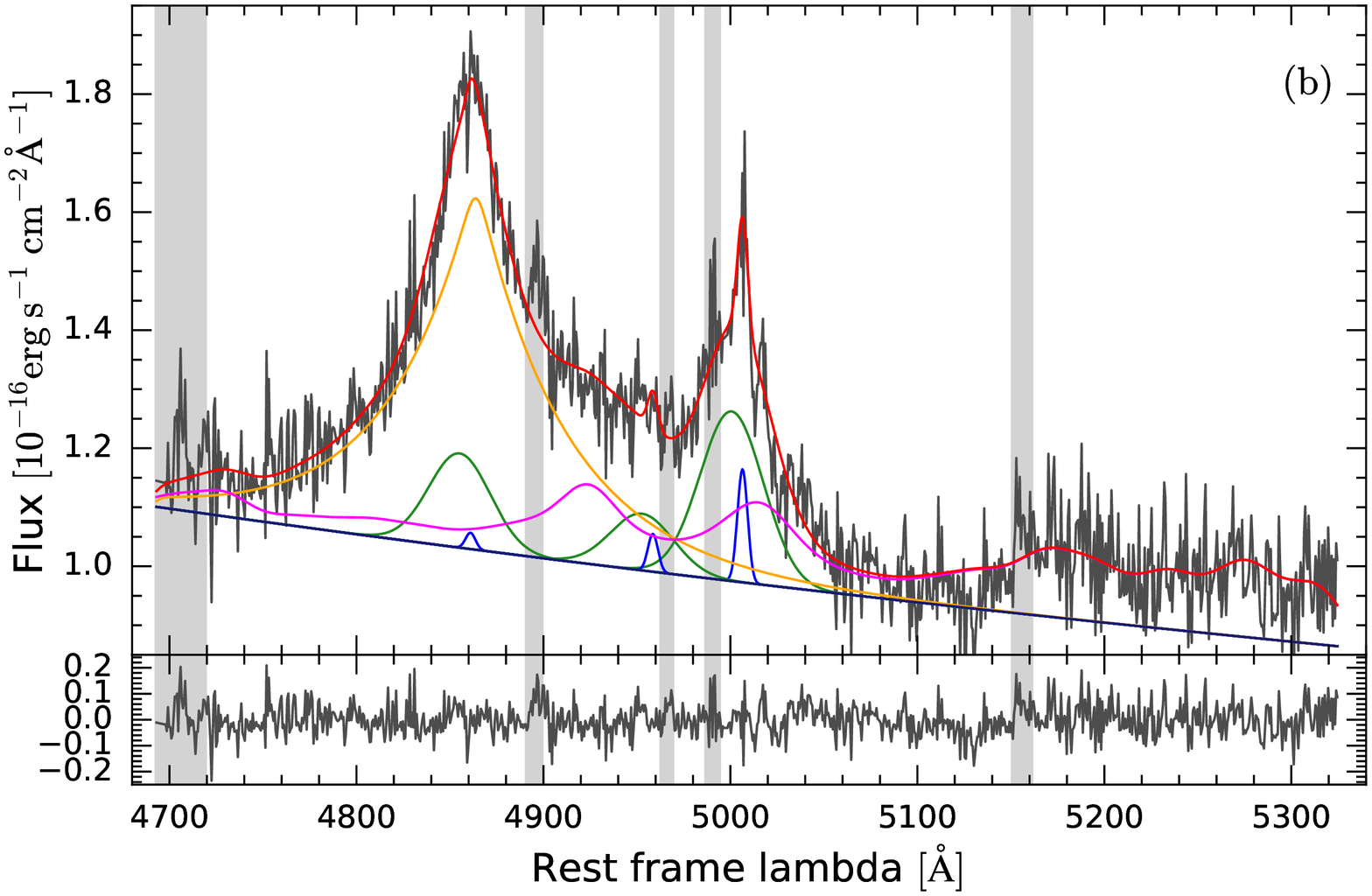}
	}
	\makebox[1\textwidth]{
		\includegraphics[width=0.5\linewidth]{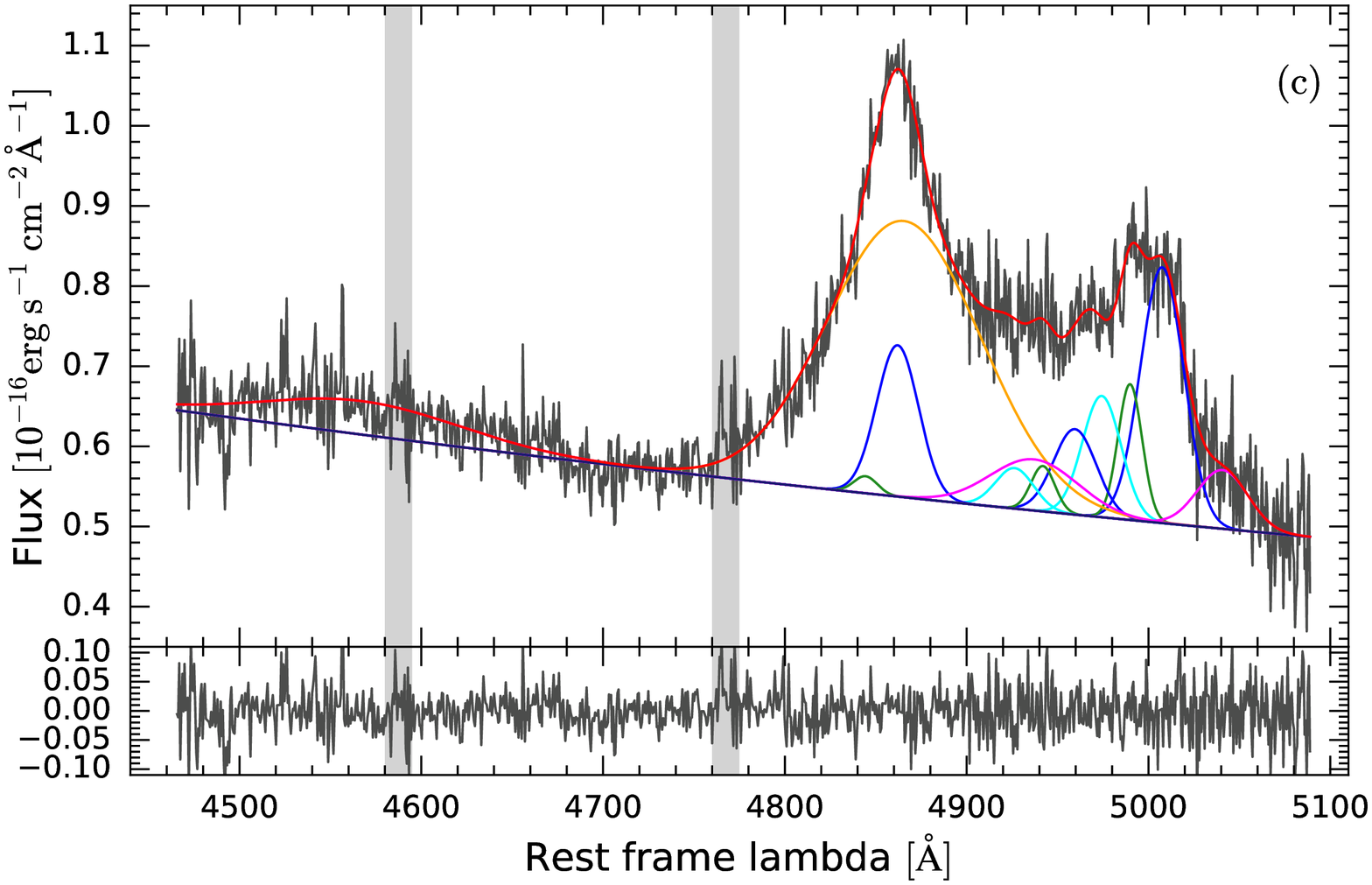}
		\includegraphics[width=0.5\linewidth]{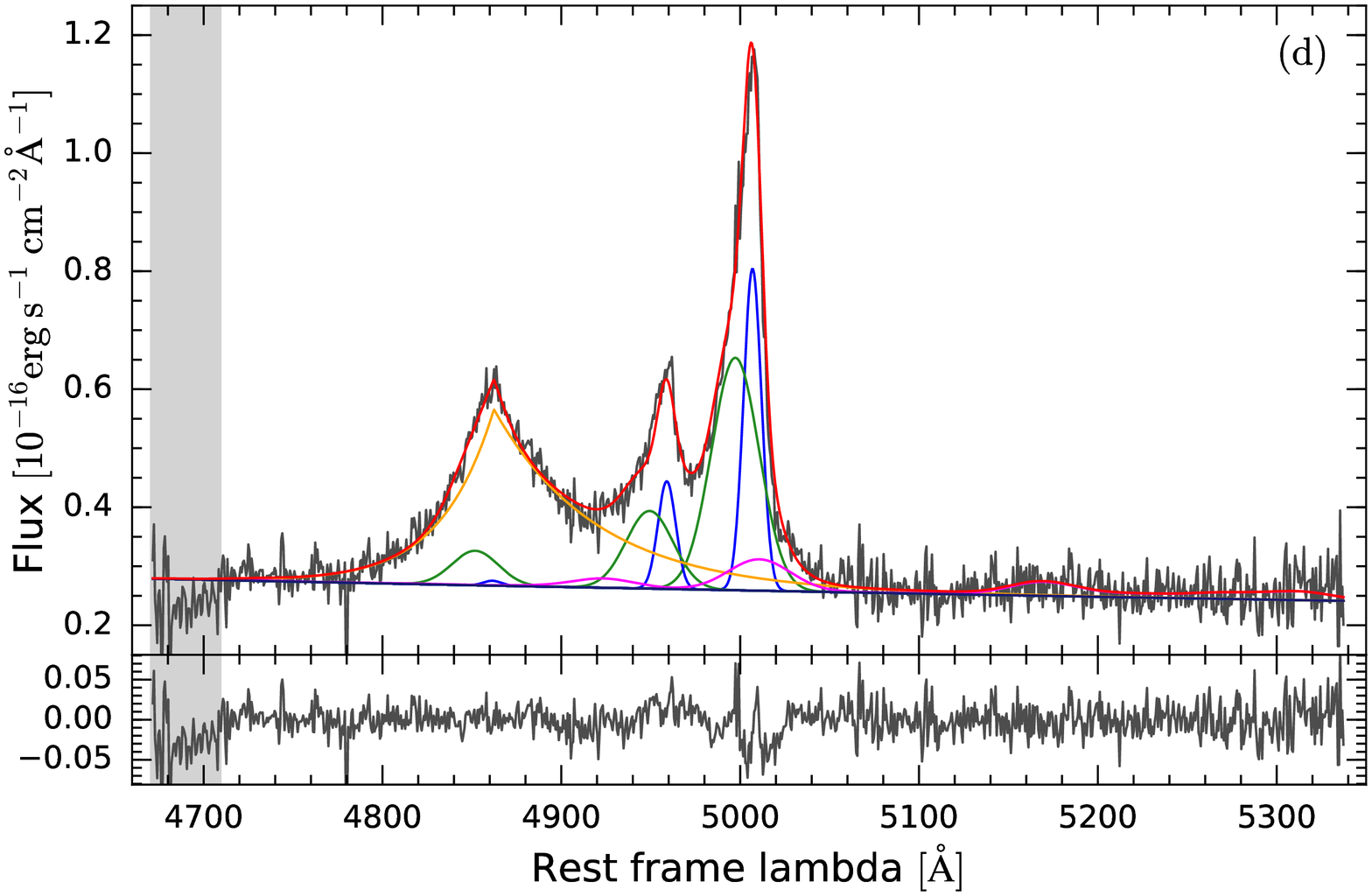}
	}
	\includegraphics[width=0.5\linewidth]{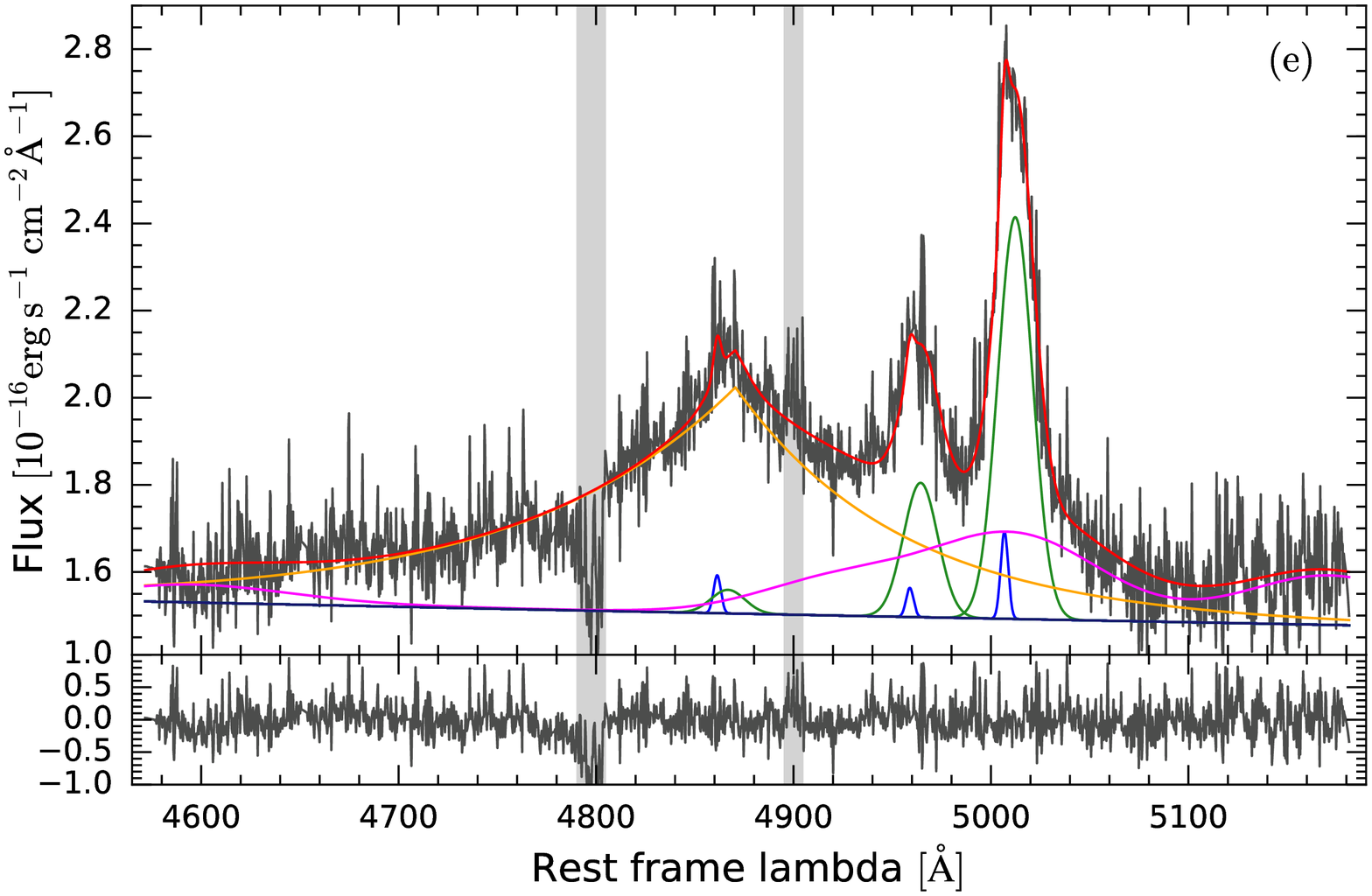}
	\label{fig:best}
\end{figure*}

Given the complexity of the spectra (see Fig. \ref{fig:best}), data analysis have been performed by using IDL custom processing scripts.
In order to reproduce the observed spectral features, we have developed a suite of models based on the IDL package 
MPFIT \citep{Markwardt2009}. Continuum and emission lines (H$\beta$, [OIII] and FeII) have been simultaneously fitted by minimizing the $\chi^2$. The statistical noise associated to the LUCI1 spectra is small (as WISSH quasars are near-IR bright sources) with respect to the systematics, e.g. the dispersion error associated to the grism. Therefore, we have estimated the statistical noise from the line-free continuum emission and have assumed it to be constant over the entire spectral range.
We have tested different spectral models, in order to find the most appropriate description of the spectrum for each WISSH quasar.
Specifically, all the spectra have been fitted with the following models:

\begin{description}
	\item[\textbf{Model A},] which consists of one Gaussian profile for fitting each component of the [OIII]$\lambda4959,5007$ \AA\ doublet; one Gaussian profile for fitting the H$\beta$ emission from the narrow line region (NLR); one broken power-law, convolved with a Gaussian curve, for fitting the broad line region (BLR) H$\beta$ emission \citep[e.g.][]{NagaoMarconiMaiolino2006,CarnianiMarconiMaiolinoEtAl2015}.
	\item[\textbf{Model B1},] which includes one Gaussian profile for fitting each component of the narrow, systemic [OIII] doublet and one Gaussian profile to account for the NLR H$\beta$ emission; one Gaussian profile for fitting each component of the broad [OIII] doublet, associated with the outflowing gas, and one Gaussian profile for fitting the corresponding broad H$\beta$ emission; a broad Gaussian component for fitting the BLR H$\beta$ emission. 
	\item[\textbf{Model B2},] which is similar to Model B1, except for a broken power-law profile, convolved with a Gaussian curve, included to describe the BLR H$\beta$ emission.
\end{description}

For both narrow and broad emission, the separation between each component of the [OIII]$\lambda4959$, $5007$ \AA\ doublet
has been fixed to  48 \AA\ in the rest-frame,  with a  ratio  of their normalizations equal to  1:3. Moreover, the standard deviation $(\sigma)$ of the two doublet components has been forced to be the same.
For narrow(broad) H$\beta$ emission,  the separation with respect to the narrow(broad) [OIII]$\lambda5007$ \AA\ emission has been fixed to  146 \AA\ in the rest-frame (assuming the same $\sigma$ of the [OIII] emission).
All models include a power-law continuum component,  estimated from features-free spectral regions at both sides of the $\rm H\beta-[OIII]$  emission lines (except for \quattro, in which the spectral region bluewards the H$\beta$ is not observed) sufficiently wide (from $\sim200$ \AA\ to $\sim300$ \AA\ in the rest-frame in case of \tre\ and \uno, respectively) to allow a meaningful constraint, and FeII emission. 

Specifically, the complex FeII emission in the spectra has been taken into account by adding to the fit a specific template based on 
observational FeII emission templates  \citep{BorosonGreen1992,Veron-CettyJollyVeron2004,TsuzukiKawaraYoshiiEtAl2006} or synthetic FeII spectral templates created by the Cloudy  plasma simulation code \citep{FerlandPorterVanHoof2013}.
Furthermore, by convolving  these templates with a Gaussian function with a $\sigma$ value ranging  from 1000 to 4000 \kms, we have created a library of  FeII emission templates for each quasar. 
In this way, we are able to describe the FeII emission in all our targets (see Fig. \ref{fig:best}) but \tre, for which we cannot find an appropriate FeII template.
Accordingly, we have decided to include in the fitting models of \tre\ three broad Gaussian components to account for the strong emission features observed at $\sim$ 4500, 4940 and 5040 \AA, respectively. We note that the widely used  \cite{BorosonGreen1992} template fails to provide a
satisfactory fit to the FeII emission in all spectra.

In case of \tre, A and B models are not able to account for the plateau-like emission feature in the $4900-4970$ \AA\ range (see Fig. \ref{fig:best}c).
Accordingly, the LUCI1 spectrum of this source have been also fitted by the following models:
\begin{description}
\item[\textbf{Model C1},] which is similar to Model B2, but also includes an additional Gaussian profile to account for an extra broad  component of [OIII]$\lambda5007$ \AA, [OIII]$\lambda4959$ \AA\ and H$\beta$ emissions.
\item[\textbf{Model C2:},] which is similar to Model C1, but includes a  broken power-law component, convolved with a Gaussian curve, to account for  the BLR H$\beta$ emission.
\end{description}

\subsection{Extinction estimate}\label{sec:sed}

The spectral energy distributions (SEDs) of the  WISSH quasars have been derived by fitting the near-UV to mid-IR photometric data (listed in Table \ref{tab:mag}) with a set of SED models created from the Type 1 AGN SED templates of \cite{RichardsLacyStorrie-LombardiEtAl2006} and applying a SMC extinction law \citep{PrevotLequeuxPrevotEtAl1984} with colour excess $\rm E(B-V)$ as free parameter. The detailed description of this procedure and the corresponding results for the entire WISSH quasars sample will be provided in a forthcoming paper (Duras et al.). 
The SMC-like extinction curve has been used as usually done for mildly reddened quasars up to z $\sim2.5$ \citep[e.g.][]{RichardsHallVandenBerkEtAl2003,HopkinsStraussHallEtAl2004}. 
Nonetheless, an additional SED fitting has been tested by adopting the mean extinction curve (MEC) from \cite{GalleraniMaiolinoJuarezEtAl2010}, which has been found to be more representative of high-z quasars ($z \sim4-6$).  
The MEC extinction law has been derived for $\lambda<5000$ \AA\  \citep{GalleraniMaiolinoJuarezEtAl2010}. For $\lambda>5000$ \AA, the dust extinction has been accounted for by considering the curve from \cite{CalzettiArmusBohlinEtAl2000}, which is also widely used to estimate the reddening in high-z objects \citep[e.g.][for the COSMOS sample]{LussoComastriSimmonsEtAl2012}.
The $\rm E(B-V)$ values derived by using the SMC or MEC$+$Calzetti extinction curve have been found to be consistent within one standard deviation for all sources. Accordingly, the extinction level does not strongly depend on the adopted extinction curve, and we consider the SMC based values as reliable for our targets.

We find that \quattro\ and \cinque\ are compatible with no extinction while, for the remaining objects, we have verified that applying the SED-based $\rm E(B-V)$ values to the typical intrinsic Type 1 quasar spectrum, provided by \cite{VandenBerkRichardsBauer2001}, we are able to reproduce  the observed LUCI1 and SDSS spectrum of each quasar.
The spectra of \tre\ and \due\ are successfully reproduced with $\rm E(B-V)=0.005$ and 0.148, respectively. In case of \uno, the spectral template modified by the SED-based $\rm E(B-V)=0.296$
fails to describe the  SDSS spectrum at $\lambda<1200$ \AA\ (rest frame).
Specifically,  the template rapidly drops at shorter wavelengths, while the observed spectrum remains almost flat down to $\sim800$ \AA.
There are two possible explanations for this: (i)
the SMC extinction curve is not appropriate to describe the extinction for this quasar; (ii)
a contamination from the host galaxy emission can be present, in addition to the quasar one, producing an excess in this wavelength range.

The application of the extinction curves from \cite{CalzettiArmusBohlinEtAl2000} and \cite{GalleraniMaiolinoJuarezEtAl2010} have not successfully reproduced the observed spectrum. We have therefore added in the SED fitting model a host galaxy component chosen among the templates of \cite{BruzualCharlot2003}, obtaining an excellent description of the data. Accordingly, we assume the SED-based $\rm E(B-V)=0.296$ for \uno.

\begin{table*}[p]
	
	\centering
	\caption{Spectral fit results derived from the different models (see Sect. \ref{sec:spectral}) applied to \textit{(a)} SDSS \quattro, \textit{(b)} \cinque, \textit{(c)} \tre, \textit{(d)} \uno\ and \textit{(e)} \due. Boldface indicates the best fit model. The spectroscopic redshift for each WISSH quasar, estimated from the best fit of LUCI1 data, is also shown.}
	\label{tab:comp}
	\footnotesize
	\subtable[SDSS J0745$+$4734\hspace{0.2cm} $\rm z_{spec}=3.225\pm0.001$]{
		\makebox[\textwidth]{
			\begin{tabular}{lcccr}
				\hline
				Parameter & Mod A & Mod B1 & \bf{Mod B2} & Units\\
				\hline
				
				$\rm \chi^{2}_{ \nu}$ &    1.46       &    1.30     &     1.27   &    \\
				FWHM$\rm ^{core}_{[OIII]}$      &    1510$\pm160$       & 440$\pm150$  & 470$\pm170$ & \footnotesize{km s$^{-1}$}    \\
				FWHM$\rm ^{broad}_{[OIII]}$      &    $-$     &   1740$\pm170$     &   1630$\pm180$   &   \footnotesize{km s$^{-1}$}  \\
				$\rm \lambda^{broad}_{[OIII]}$    &     $-$        &    4999$\pm1$     &    4999$\pm1$  & \footnotesize{\AA}    \\
				Flux [OIII]$\rm ^{broad}$/ [OIII]$\rm ^{core}$    &     $-$      &    8.5$\pm1.6$     &   6.8$\pm1.3$   &    \\
				$\rm \lambda^{BLR}_{H\beta}$      &    4866$\pm2$   &  4867$\pm3$     &   4862$\pm1$ &  \footnotesize{\AA}       \\
				FWHM$\rm ^{BLR}_{H\beta}$       &    6920$\pm210$   &    8130$\pm230$   &   8600$_{-200}^{+230}$  & \footnotesize{km s$^{-1}$}    \\
				Flux H$\rm \beta^{BLR}$/H$\rm \beta^{Tot}$    &   0.99$_{-0.05}^{+0.01}$    &    0.93$\pm$0.06          &   0.98$_{-0.08}^{+0.02}$  &      \\
				\hline
				
			\end{tabular}
			
		}
	}
	\subtable[SDSS J0900$+$4215\hspace{0.2cm} $\rm z_{spec}=3.294\pm0.001$]{
		\makebox[\textwidth]{
			\begin{tabular}{lcccr}
				\hline
				Parameter & Mod A & Mod B1 & \bf{Mod B2} & Units\\
				\hline
				
				$\rm \chi^{2}_{ \nu}$ &    1.95       &    1.65     &     1.57   &    \\
				FWHM$\rm ^{core}_{[OIII]}$      &    2030$\pm180$       & 250$\pm130$  & 300$\pm130$ & \footnotesize{km s$^{-1}$}    \\
				FWHM$\rm ^{broad}_{[OIII]}$      &    $-$     &   2340$\pm270$     &   2240$\pm170$   &   \footnotesize{km s$^{-1}$}  \\
				$\rm \lambda^{broad}_{[OIII]}$    &     $-$        &    5004$\pm3$     &    4999$\pm1$  & \footnotesize{\AA}    \\
				Flux [OIII]$\rm ^{broad}$/ [OIII]$\rm ^{core}$    &     $-$      &    19.1$\pm3.3$     &   11.7$\pm1.8$   &    \\
				$\rm \lambda^{BLR}_{H\beta}$      &    4867$\pm2$   &  4871$\pm3$     &   4861$\pm1$ &  \footnotesize{\AA}       \\
				FWHM$\rm ^{BLR}_{H\beta}$       &    3740$\pm210$   &    6900$\pm280$   &   3210$\pm190$  & \footnotesize{km s$^{-1}$}    \\
				Flux H$\rm \beta^{BLR}$/H$\rm \beta^{Tot}$    &   0.93$\pm0.05$    &  0.92$_{-0.06}^{+0.03}$            &   0.98$\pm$0.02  &      \\
				\hline
				
			\end{tabular}
			
		}
	}
	\subtable[SDSS J1201$+$1206\hspace{0.2cm} $\rm z_{spec}=3.512\pm0.002$]{
		\makebox[\textwidth]{
			\begin{tabular}{lcccr}
				\hline
				Parameter & Mod A & \bf{Mod C1} & Mod C2 & Units\\
				\hline
				
				$\rm \chi^{2}_{ \nu}$ &    1.41       &    1.26     &     1.30   &    \\
				FWHM$\rm ^{broad, 5007 \AA}_{[OIII]}$      &    $($2560$\pm200$       & 1670$\pm190$  & 1700$\pm200)^a$ & \footnotesize{km s$^{-1}$}    \\
				FWHM$\rm ^{broad, shifted}_{[OIII]}$      &    $-$     &   940$\pm210$     &   960$\pm180$   &   \footnotesize{km s$^{-1}$}  \\
				& & $($1510$_{-180}^{+300}$ &  1700$\pm280)^b$ & \\
				$\rm \lambda^{broad, shifted}_{[OIII]}$    &     $-$        &    4990$\pm2$     &    4989$\pm2$  & \footnotesize{\AA}    \\
				& &  $($4970$\pm1$ & 4973$\pm2)^b$ & \\
				$\rm \lambda^{BLR}_{H\beta}$      &    4867$\pm3$   &  4865$\pm2$     &   4858$\pm2$ &  \footnotesize{\AA}       \\
				FWHM$\rm ^{BLR}_{H\beta}$       &    7470$\pm280$   &    6160$\pm250$   &   4800$\pm200$  & \footnotesize{km s$^{-1}$}    \\
				Flux H$\rm \beta^{BLR}$/H$\rm \beta^{Tot}$    &   0.92$_{-0.10}^{+0.08}$    &    0.86$\pm$0.07          &   0.96$_{-0.1}^{+0.04}$  &      \\
				\hline
				
			\end{tabular}
			
		}
		
	}
	\subtable[SDSS J1326$-$0005\hspace{0.2cm} $\rm z_{spec}=3.303\pm0.001$]{
		\makebox[\textwidth]{
			\begin{tabular}{lcccr}
				\hline
				Parameter   & Mod A & Mod B1 & \bf{Mod B2} & Units\\
				\hline
				
				$\rm \chi^{2}_{ \nu}$ &     2.42        &    1.81    &    1.51  &    \\
				FWHM$\rm ^{core}_{[OIII]}$      &    1290$\pm160$        &  730$\pm160$   &  710$\pm170$  & \footnotesize{km s$^{-1}$}   \\
				FWHM$\rm ^{broad}_{[OIII]}$        &    $-$     &   1980$\pm170$     &   1870$\pm170$   & \footnotesize{km s$^{-1}$}     \\
				$\rm \lambda^{broad}_{[OIII]}$     &      $-$        &    4997$\pm1$     &  4997$\pm1$  & \footnotesize{\AA}    \\
				Flux [OIII]$\rm ^{broad}$/ [OIII]$\rm ^{core}$    &      $-$      &    1.9$\pm1.0$     & 1.9$\pm0.9$   &    \\
				$\rm \lambda^{BLR}_{H\beta}$      &    4865$_{-3}^{+1}$     &  4874$\pm2$     &  4862$\pm1$  & \footnotesize{\AA}       \\
				FWHM$\rm ^{BLR}_{H\beta}$      &    5480$_{-200}^{+250}$   &  5520$_{-240}^{+180}$      &  3700$\pm160$  & \footnotesize{km s$^{-1}$}    \\
				Flux H$\rm \beta^{BLR}$/H$\rm \beta^{Tot}$ &   0.92$\pm0.06$       &    0.8$\pm$0.1          &   0.93$_{-0.09}^{+0.07}$  &      \\

				\hline
			\end{tabular}
		}
	}

	\subtable[SDSS J1549$+$1245\hspace{0.2cm} $\rm z_{spec}=2.365\pm0.001$]{
		\makebox[\textwidth]{
			\begin{tabular}{lcccr}
				\hline
				Parameter  & Mod A & Mod B1 & \bf{Mod B2} & Units\\
				\hline
				$\rm \chi^{2}_{ \nu}$   &    1.45      &    1.42     &     1.37   &    \\
				FWHM$\rm ^{core}_{[OIII]}$      &     1260$\pm210$        & 280$\pm160$  &  270$\pm160$ & \footnotesize{km s$^{-1}$}    \\
				FWHM$\rm ^{broad}_{[OIII]}$       &    $-$     &   1250$\pm210$    &    1240$\pm210$   & \footnotesize{km s$^{-1}$}    \\
				$\rm \lambda^{broad}_{[OIII]}$     &     $-$        &    5012$\pm1$    &   5012$\pm1$  & \footnotesize{\AA}    \\
				Flux [OIII]$\rm ^{broad}$/ [OIII]$\rm ^{core}$    &    $-$      &    20$\pm$2     &   21$\pm3$   &    \\
				$\rm \lambda^{BLR}_{H\beta}$      &    4864$\pm2$   &  4882$\pm2$     &   4871$\pm2$ & \footnotesize{\AA}       \\
				FWHM$\rm ^{BLR}_{H\beta}$      &    7810$\pm250$   &    11100$_{-260}^{+310}$    &   8340$\pm280$  & \footnotesize{km s$^{-1}$}    \\
				Flux. H$\rm \beta^{BLR}$/H$\rm \beta^{Tot}$  &   0.98$_{-0.10}^{+0.02}$    &    0.94$_{-0.09}^{+0.06}$     &   0.98$_{-0.09}^{+0.02}$  &      \\
				
				\hline
			\end{tabular}
		}
	} 
	\flushleft{$^a$ Values derived for the broad [OIII] component at 5007 \AA\ (see Sect. \ref{sec:best}). \\ 
	\noindent $^b$ Values derived for the additional broad, shifted [OIII] component included in models C1 and C2 (see Sect. \ref{sec:spectral}).}
	\vspace{0.3cm}
\end{table*}


\section{Results}\label{sec:results}

\subsection{Best-fit Models}\label{sec:best}

The rest-frame main spectral parameters, derived from the different models applied to the LUCI1 data, are shown in Table \ref{tab:comp}. In the following, the quoted errors refer to one standard deviation confidence level.
The FWHM values are not corrected for instrumental broadening, as they are much broader than the  LUCI1 spectral resolution (rest-frame FWHM$_{\rm ins}$  $\lesssim 20$ \kms). 

\begin{description}

\item[\textbf{SDSS J0745$+$4734:}] the LUCI1 spectrum of this WISSH quasar exhibits a prominent, well defined [OIII] doublet emission emerging from the extended red wing of the H$\beta$ profile, as shown in Fig. \ref{fig:best}a. The best-fit description of this spectrum is provided by  Model B2, with an associated $\chi^{2}_{ \nu}$ = 1.27.

It is possible to accurately constrain the weak FeII emission underlying the [OIII] region thanks to the spectral coverage extended up to $\sim5400$ \AA. Two  Gaussian components are necessary to properly reproduce the skewed [OIII] emission profile, dominated by a blueshifted, broad (i.e., FWHM$_{\rm [OIII]}^{\rm broad}\sim1630$ \kms) component. The ratio between the integrated flux of the broad [OIII] component, centered at $\sim$ 4999 \AA, and the core (FWHM$_{\rm [OIII]}^{\rm core}\sim470$ \kms) component is $\sim6.8$. The H$\beta$  emission is characterised by a very large profile, with a FWHM$_{\rm H\beta}^{\rm BLR}\sim8600$ \kms, i.e. the highest value derived within our sample.
\vspace{0.3cm}

\item[\textbf{SDSS J0900$+$4215:}] the LUCI1 spectrum of this quasar is dominated by the H$\beta$ emission and shows a strong FeII pseudo-continuum (Fig. \ref{fig:best}b). Model B2 yields the best-fit for this spectrum ($\chi^{2}_{ \nu}$ = 1.57).

The [OIII] emission is best-fitted by two Gaussian components, i.e.  a narrow (FWHM$_{\rm [OIII]}^{\rm core}\sim300$ \kms) one representing the systemic emission, and an extremely broad (FWHM$_{\rm [OIII]}^{\rm broad}\sim2240$ \kms) and blueshifted component with centroid $\lambda_{\rm OIII}^{\rm broad}\sim4999$ \AA.
The broad [OIII] component dominates the emission with a  [OIII]$^{\rm broad}$/ [OIII]$^{\rm core}$ flux ratio of $\sim12$. The skewed H$\beta$ profile is better reproduced by a broken power-law. We derive a FWHM$_{\rm H\beta}^{\rm BLR}\sim3200$ \kms, which is the smallest value in our sample.
\vspace{0.3cm}

\item[\textbf{SDSS J1201$+$1206:}] the most striking feature in the LUCI1 spectrum of this quasar is  the \textit{plateau}-like emission ranging redwards the peak of the strong H$\beta$ emission up to the peak of the [OIII]$\lambda5007$ \AA, that Models A and B are not able to account for (see Fig. \ref{fig:best}c and Sect. \ref{sec:spectral}).
We have therefore applied to the data models C1 and C2, which include three Gaussian profiles for each component of the [OIII] doublet, in order to reproduce the broad/complex [OIII] emission. Model C1 yields the best description of the spectrum, with an associated $\chi^{2}_{ \nu}$ = 1.26 (see Table \ref{tab:comp}).

The  bulk of the [OIII] emission comes from a broad Gaussian line centered at 5007 \AA, with a FWHM of $\sim$ 1700 \kms. Such a large value clearly indicates that this line cannot be ascribed to the NLR, but it is associated with high-velocity, outflowing gas, and suggests degeneracy between the Gaussian profiles used in the model to fit the [OIII] emission.
The H$\beta$  emission is best-fitted by the combination of three Gaussian profiles. The FWHM of the BLR H$\beta$ line results $\sim$ 6200 \kms.
J1201$+$1206 is also characterised by emission features likely associated to FeII, being relevant in the region bluewards the H$\beta$ and in correspondence of the \textit{plateau}-like feature.
\vspace{0.3cm}

\item[\textbf{SDSS J1326-0005:}] the LUCI1 spectrum of this WISSH quasar (see Fig. \ref{fig:best}d)  resembles that typically observed for less luminous quasars, with a very prominent [OIII] emission with respect to the H$\beta$ one.  Model B2 provides a good description of the LUCI1 spectrum, with an associated reduced $\chi^{2}_{ \nu}$ =  1.51.

Two Gaussian components are needed to properly reproduce the broad and skewed profile of the [OIII] emission.
The core component shows a FWHM$^{\rm core}_{\rm [OIII]}$  of $\sim$ 700 \kms, while the broad component is centered at 4997 $\pm$ 1 \AA, with  a FWHM$^{\rm broad}_{\rm [OIII]}$ $\sim$ 1900 \kms. The ratio between the flux of  broad and core component is $\sim$ 1.9.  
The H$\beta$  emission is dominated by the contribution from the BLR and, because of the highly skewed and asymmetric profile, cannot be reproduced by a Gaussian function and requires the application of  a broken power-law component, for which we measure a FWHM$^{\rm BLR}_{\rm H\beta}\sim$ 3700 \kms.
\vspace{0.3cm}

\item[\textbf{SDSS J1549$+$1245:}] the $\rm H$-band spectrum of \due\ is  characterised by a very broad  (FWHM$_{\rm H\beta}^{\rm BLR}\sim8300$ \kms) H$\beta$ profile (see Fig.\ref{fig:best}e), similarly to the \quattro\ spectrum. The [OIII] doublet clearly emerges from the very extended H$\beta$ red wing.   Model B2 provides the best-fit model for this quasar ($\chi^{2}_{ \nu}$ = 1.37).

The prominent  H$\beta$ line makes difficult an accurate fitting of the  [OIII] emission, as part of it (in particular the broad wings) is likely blended in the H$\beta$ profile (similarly to the underlying FeII emission).
Nonetheless,   two  Gaussian components are necessary to properly reproduce the [OIII] emission, with the broad one (i.e. FWHM$^{\rm broad}_{\rm [OIII]}\sim$1240 \kms) much stronger than the core emission. The centroid of the broad Gaussian is at 5012 $\pm$ 1 \AA, indicating  that the bulk of the [OIII] emission is associated to  receding material.
\vspace{0.3cm}

\end{description}
\noindent Table \ref{tab:comp} lists the spectroscopic redshift of our sources derived from the best-fit model. Our estimates significantly $(>3\sigma)$ differ from the SDSS redshifts (see Table \ref{tab:mag}), which are based on UV emission lines and derived by an automatic procedure. \cinque\ is the only source for which we find a $z_{\rm spec}$ value comparable to $z_{\rm SDSS}$. 

We have derived an additional redshift estimate of our targets from the CIV line detected in the SDSS spectra. The CIV-based redshifts are systematically lower (with blueshifts $\lesssim 800$ \kms) than those derived from the H$\beta$ line, except for \uno, which does not show any significant shift (details on the CIV emission line properties and blueshifts in WISSH quasars will be discussed in a forthcoming paper by Vietri et al.).

\subsection{2D and "Off-nuclear" Spectra}\label{sec:2d}

\begin{figure*}[]
	\centering
	\caption{\textit{(a)} 2D LUCI1 spectrum of \tre. Blue to red colors indicate increasing counts. Since the LUCI1 spectra have been acquired with PA = 0 deg, top and bottom correspond to North and South directions, respectively. The black solid lines indicate the apertures used for extracting the off-nuclear spectra. 
	Panels \textit{(b)} and \textit{(c)} show the "off-nuclear" spectra of J1201+1206 extracted in 1 pixel ($\sim$ 1.9 kpc) width regions at increasing distances from the peak of the spatial profile of the continuum emission. Negative(positive) pixel values correspond to South(North) direction.
    In both panels, the offset increases from top to bottom, as indicated by the corresponding pixels interval.  Spectral components are indicated as in Fig. \ref{fig:best}c. The vertical line corresponds to $\lambda=5007$ \AA, while the dashed red line indicates the best-fit model of the central spectrum extracted from the two-pixels region [-1,1], plotted with arbitrary normalization for display purpose only. Grey bands indicate the regions excluded from the fit due to the presence of telluric features.}
	\makebox[1\textwidth]{
		\subfigure[]{\includegraphics[width=0.7\linewidth]{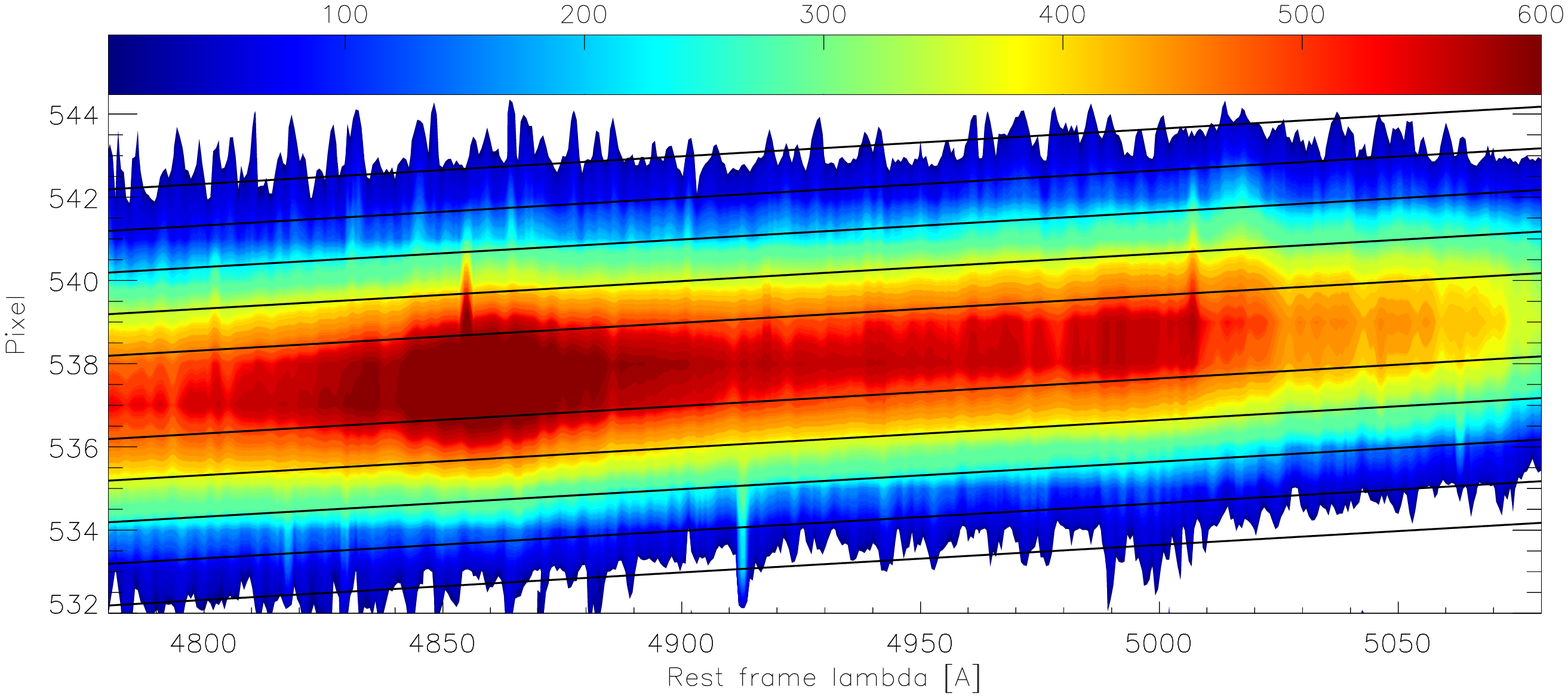}}
	}
	\makebox[1\textwidth]{
		
		\subfigure[]{
		\includegraphics[width=0.42\linewidth]{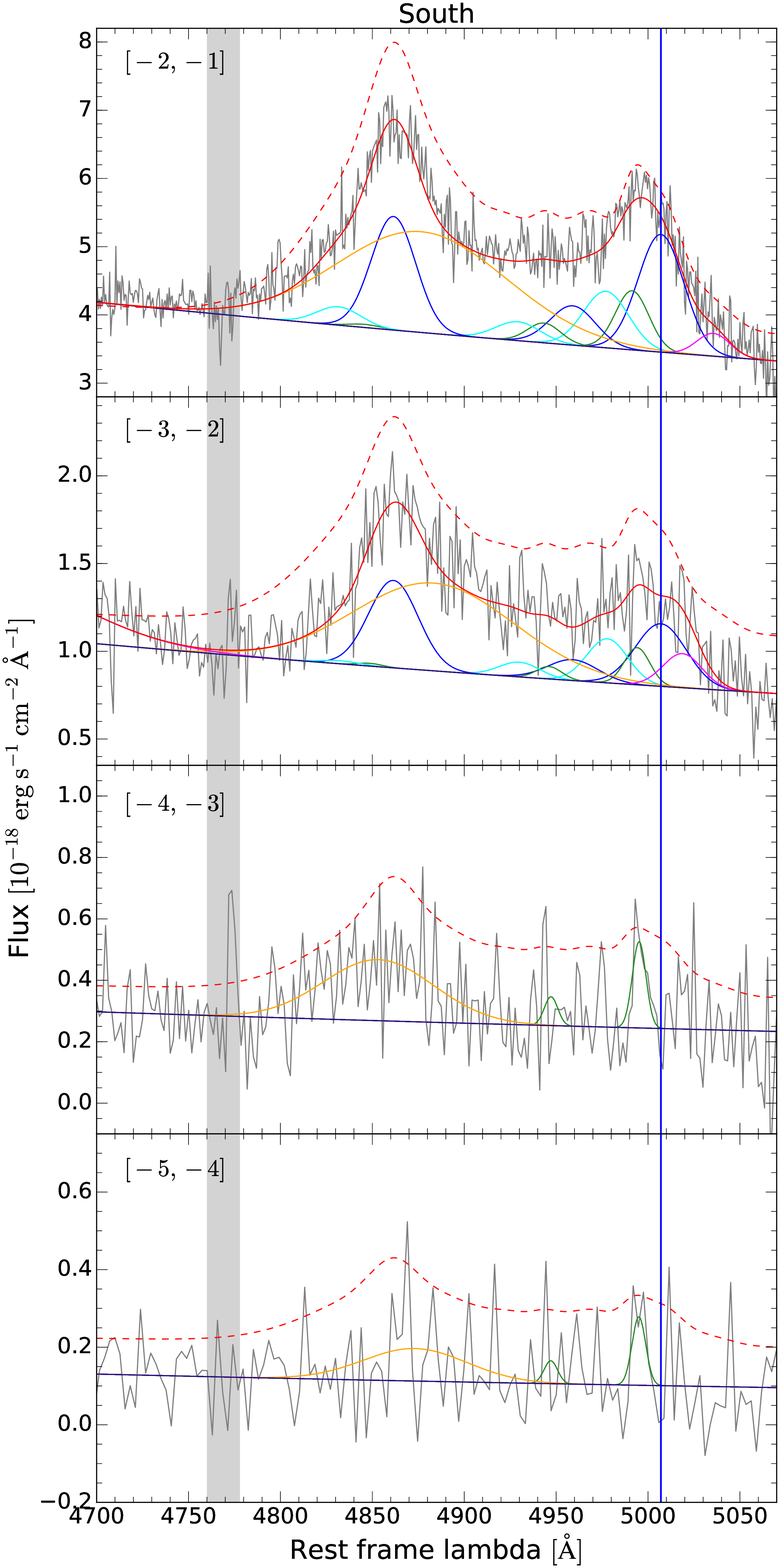}}
		\subfigure[]{
		\includegraphics[width=0.42\linewidth]{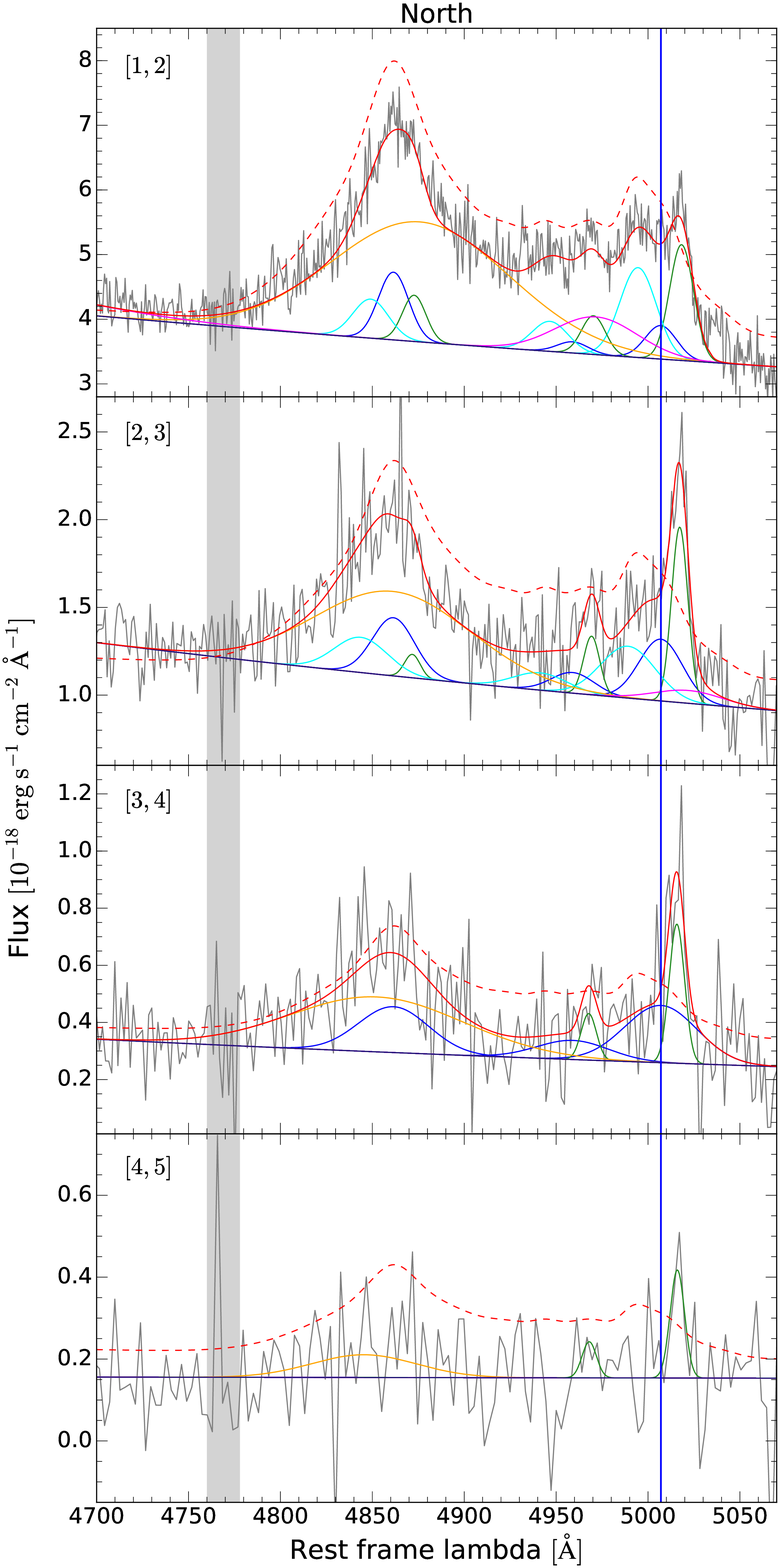}}
	}
	\label{fig:2d}
	\vspace{-0.5cm}
\end{figure*}

\begin{figure*}
	\centering
	\caption{\textit{(a)} Velocity shift of the blue/redshifted [OIII] component with respect to 5007 \AA\ (dashed line), detected in the "off-nuclear" spectra of \tre\ extracted from different apertures. Negative(positive) pixel values correspond to South(North) direction (1 pixel $\sim$ 1.9 kpc at the redshift of \tre).
	\textit{(b)} The ratio between the integrated flux of the two [OIII] components (i.e. the broad one at 5007 \AA\ and the blue/redshifted one) and the BLR H$\beta$ emission for the "off-nuclear" spectra of \tre. The yellow bands highlight the presence of extended [OIII] emission.
	\textit{(c)} Maximum velocity of the broad [OIII] component at 5007 \AA\ and the blue/redshifted [OIII] component detected in the "off-nuclear" spectra of \tre.}
	\makebox[1\textwidth]{
		\includegraphics[width=0.5\linewidth]{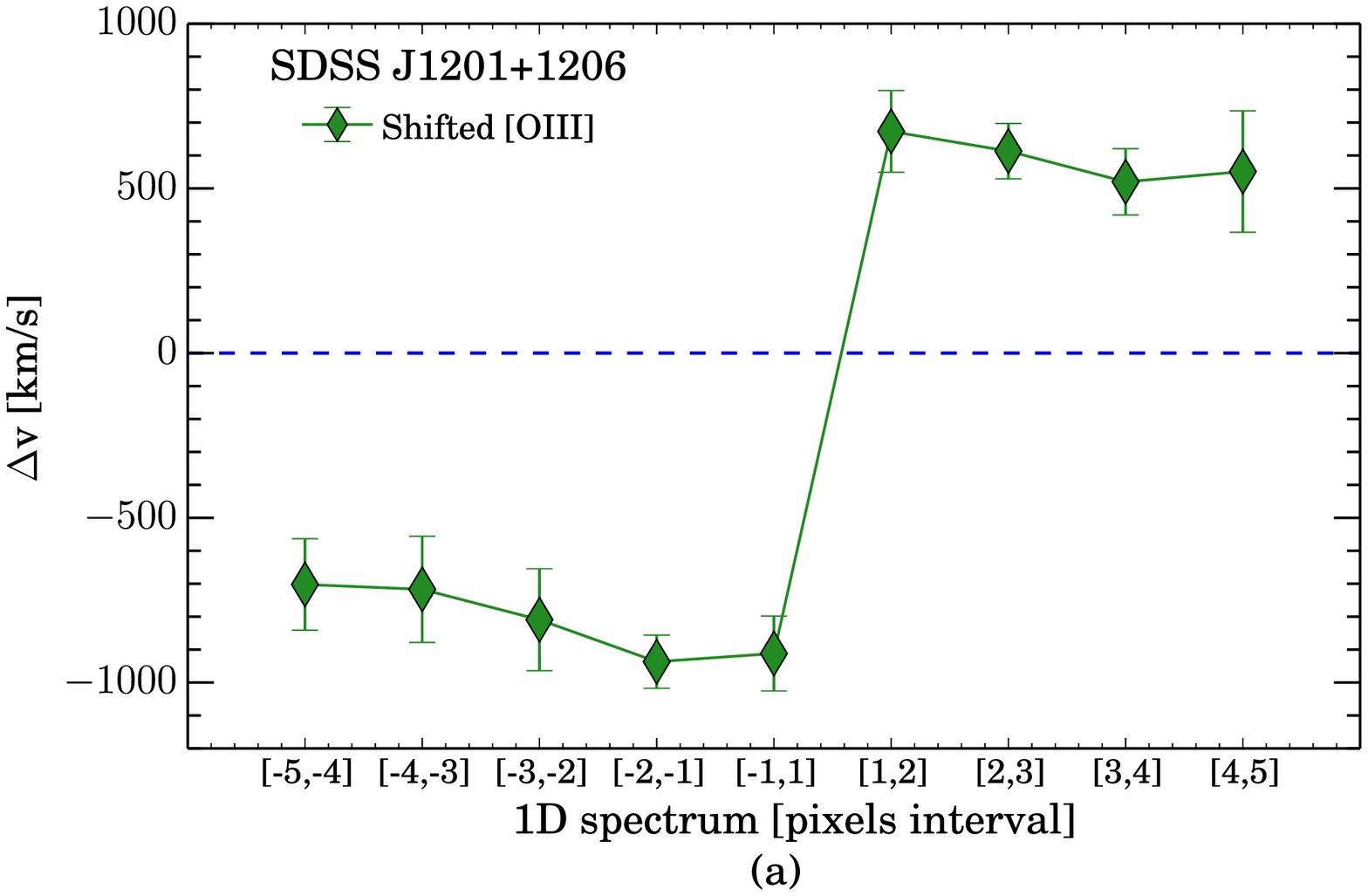}

	}
	\makebox[1\textwidth]{		

		\includegraphics[width=1\linewidth]{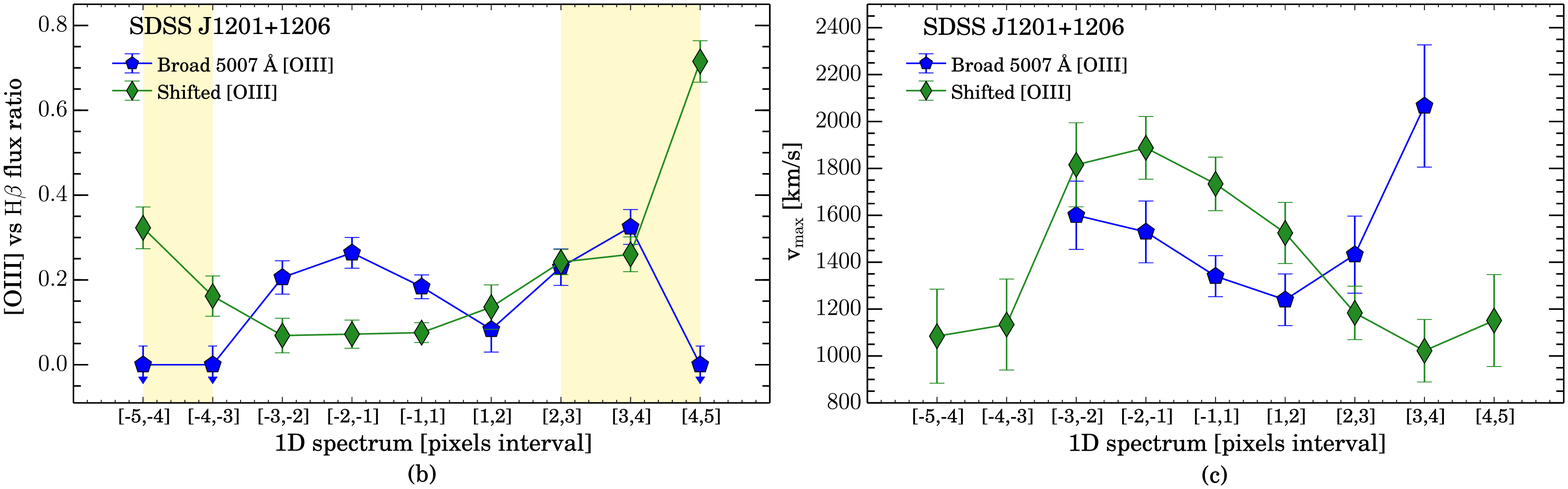}
	}
	\label{fig:offproper}
	\vspace{-0.5cm}
\end{figure*}

The presence of ionised, extended emitting gas can be investigated by extracting from the 2D spectrum, obtained from 1 arcsec slit with PA = 0 deg, 1D spectra at increasing offset distances from the central position.  In fact, going away from the center, we expect a reduced contamination from the nuclear emission and, therefore, possible spatially extended [OIII] features should be enhanced \citep{PernaBrusaCresciEtAl2015}. This approach can provide useful hints about emission components roughly oriented along the slit.

First, for each quasar,  we have extracted the spectrum from an aperture of 2 pixels (i.e. the [-1,1] spectrum of the [-1,0] and [0,1] pixels, with 1 pixel corresponding to 0.25 arcsec and $\sim$ 2 kpc at the redshifts of our targets),  centred on the peak of the spatial profile of the continuum emission. Fig. \ref{fig:2d}a shows the case for \tre, while the 2D spectra of the remaining quasars can be found in the Appendix. Negative(positive) pixel values correspond to the North(South) direction. Spectra from apertures of 1 pixel width have been then extracted for studying the "off-nuclear" spectra at increasing distance from the central peak, both in the South and North direction, as far as the spectra were not  too noisy for deriving meaningful results. For each 1D spectrum we have carried out a detailed analysis similar to that performed for the integrated spectrum (see Sect. \ref{sec:spectral}), as displayed in Fig. \ref{fig:2d} for the quasar \tre. 

Moreover, in order to trace any additional extended feature not ascribed to the nuclear emission, we have calculated the ratio between the integrated flux  of the [OIII] and BLR H$\beta$ emission for each aperture. Since the BLR H$\beta$ traces the nuclear emission, an increase of the relative strength of the [OIII] component  with respect to the BLR H$\beta$ cannot be due to the contamination from nuclear emission   and, therefore, indicates the presence of  truly extended emission.
This has highlighted evidence of possible extended [OIII] emission in three out of five sources, namely \quattro, \tre\ and \uno. In this section, we present and discuss in details the results obtained for \tre, while the results for the remaining sources are presented in the Appendix.
\vspace{0.4cm} 
 
\tre\ shows a very complex scenario, with extended [OIII] emission characterised by blueshifted and redshifted components, which are likely associated to a bipolar outflow.
The "off-nuclear" flux-calibrated spectra, extracted along South and North direction, together with their multi-component fits, are shown in Fig. \ref{fig:2d}b and \ref{fig:2d}c, respectively. As expected, the signal-to-noise ratio  significantly decreases in the outer apertures. All "off-nuclear" spectra show broad H$\beta$ emission line, indicating that none of them is  totally free from nuclear contamination. However, blueshifted and redshifted emission features, related to extended [OIII], clearly emerge in addition to the nuclear light.
Furthermore, [OIII] emission up to $\sim 5$ kpc is characterised by the presence of a broad component, centered at 5007 \AA, similarly to the profile found in the integrated spectrum (see Fig. \ref{fig:best}c and Sect. \ref{sec:best}).
 
More specifically, Fig. \ref{fig:2d}b shows the presence of an emission line at $\sim$ 4993 \AA\ along South direction, interpreted as blueshifted [OIII], and visible at distance up to $\rm \sim5-7$ kpc (i.e. in the [-4,-3] and [-5,-4] spectra). On the contrary, a prominent emission line redwards of 5007 \AA, i.e. centered at $\sim5017$ \AA, clearly emerges in all spectra from the North side up to $\sim 7$ kpc (Fig. \ref{fig:2d}c).
This redshifted  [OIII] emission component might be interpreted as the red wing of the broad (FWHM $\sim$ 1700 km s$^{-1}$) [OIII] line detected in the integrated spectrum at 5007 \AA\ (Fig. \ref{fig:best}c).
In Fig. \ref{fig:offproper}a, we plot the velocity shift $\rm \Delta v$ of the blue/redshifted [OIII] component at different apertures, with respect to 5007 \AA. The velocity shift between the blueshifted and redshifted [OIII] emission is $\sim$ 1500 \kms. Such a value cannot be explained in terms of  rotation, as it is a factor of $\gtrsim4$  larger than the typical value observed for a $z \sim 2-3$ galaxy \citep{BoucheCresciDaviesEtAl2007}. Furthermore, it would imply a very large mass enclosed within 7 kpc, i.e. $\sim4\times10^{12}$ $\rm M_{\odot}$. The bipolar outflow scenario can be therefore considered as the most likely explanation for the velocity shift observed in \tre.

Fig. \ref{fig:offproper}b displays the integrated flux ratio between the two [OIII] components (i.e. the broad one at 5007 \AA\ and the blue/redshifted one) and the BLR H$\beta$ emission as a function of the aperture. The broad [OIII] emission component is not observed at distances larger than 3 pixels from the central position, while the relative strength of the blue/redshifted component increases with distance, implying the presence of an extended outflow.
The maximum velocity of these two [OIII] components, defined as $\rm v_{max}=|\Delta v| + 2\sigma$, derived from each "off-nuclear" spectrum is plotted in Fig. \ref{fig:offproper}c, indicating that both components are associated to outflowing ($\rm v_{max}>1000$ \kms) ionised gas.\\

\begin{figure*}[h]
	\centering
	\caption{\textit{(a)} 2D LUCI1 spectrum of \quattro. Blue to red colours indicate increasing counts. Top and bottom correspond to North and South direction, respectively. The black solid lines indicate the apertures used for extracting the off-nuclear spectra. 
	\textit{(b)} The integrated flux ratio of the shifted and systemic [OIII] components with respect to the BLR H$\beta$ emission line detected in the "off-nuclear" spectra of \quattro\ extracted from different apertures. Negative(positive) pixel values correspond to South(North) direction (1 pixel $\sim$ 1.9 kpc at the redshift of \quattro).
	The yellow band highlights the presence of extended [OIII] emission.
	\textit{(c)} Maximum velocity of the shifted  and systemic [OIII] components detected in the "off-nuclear" spectra of \quattro.} 
		\vspace{0.5cm}
		\subfigure[]{\includegraphics[width=0.7\linewidth]{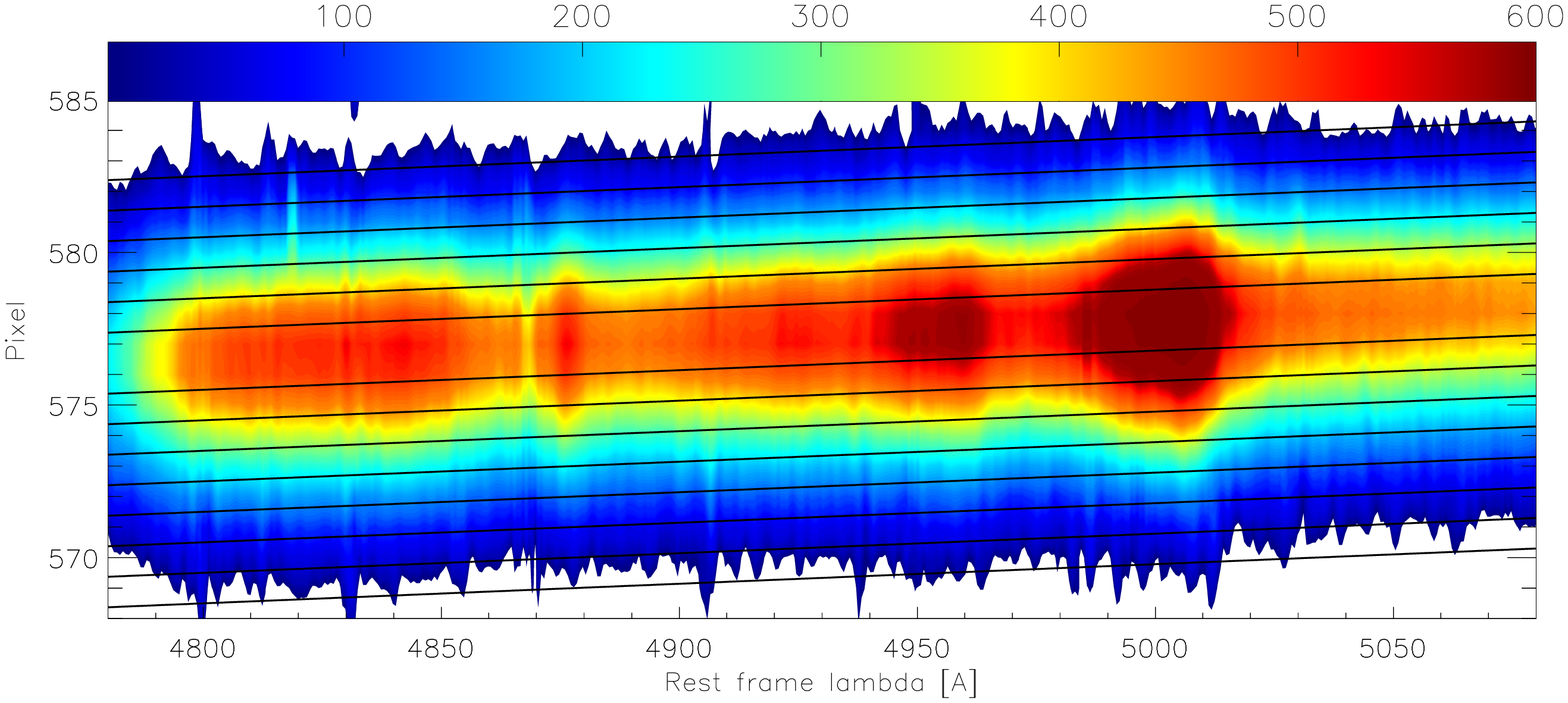}}
		\includegraphics[width=0.95\linewidth]{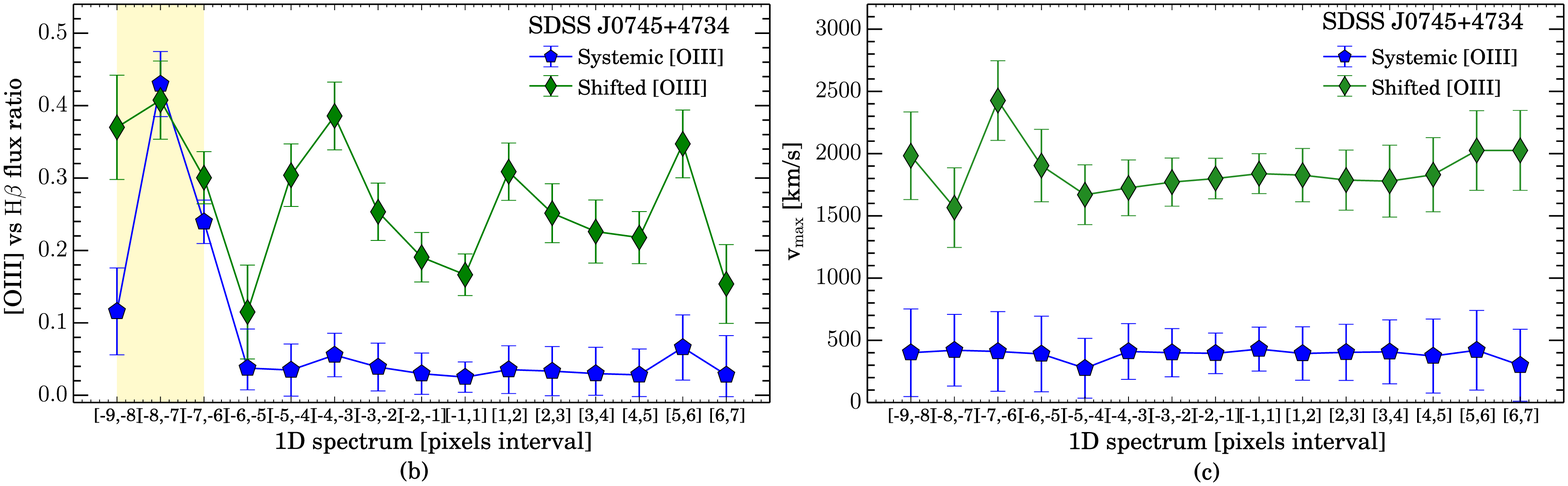}
	\label{fig:vel-flussi0745}
\end{figure*}

\begin{figure*}[]
	\centering
	\caption{Same as Fig. \ref{fig:vel-flussi0745} but for \uno\ (1 pixel $\sim$ 1.9 kpc at the redshift of this source).}
			\vspace{-0.2cm}
			\subfigure[]{\includegraphics[width=0.69\linewidth]{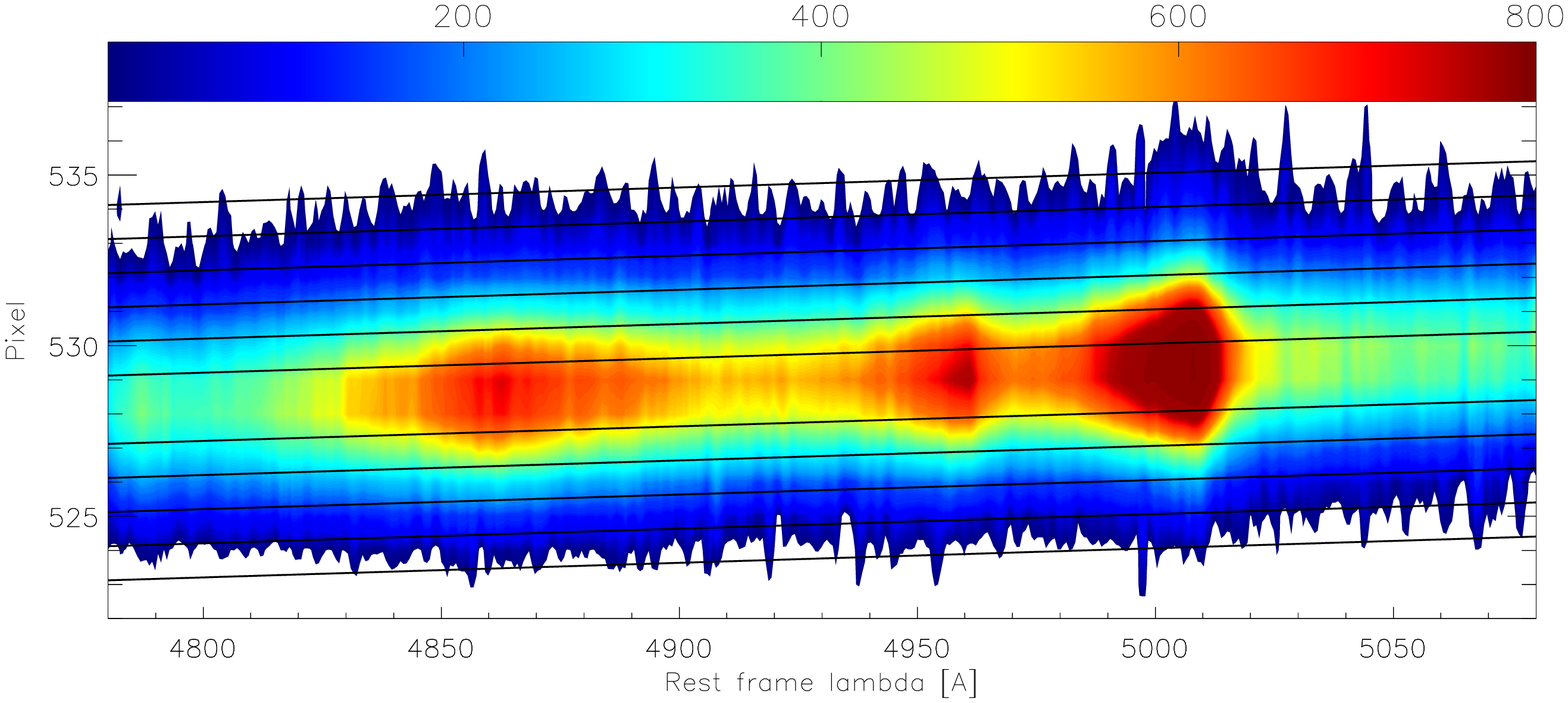}}
			\includegraphics[width=0.95\linewidth]{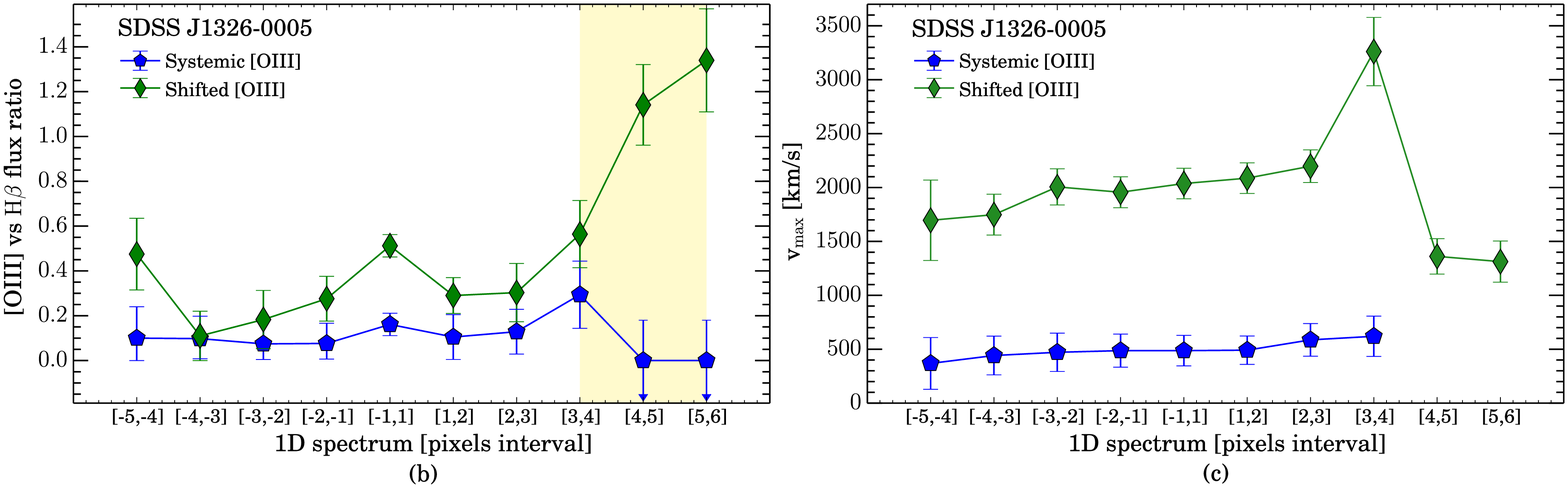}
		\vspace{-0.3cm}
	\label{fig:vel-flussi1326}
\end{figure*}

The extended [OIII] emission revealed in the other two quasars, i.e. \quattro\ and \uno, shows a different spatial behavior. As for \quattro,  we have hints of systemic, narrow [OIII] emission extended in the South direction, at a distance of 6 -- 8 pixels (i.e.  $\sim$ 11 -- 13 kpc), provided by an increase of the ratio between the integrated flux of the [OIII]$\lambda5007$ \AA\ and the BLR H$\beta$ component (see Fig. \ref{fig:vel-flussi0745}).
The analysis of the "off-nuclear" spectra of \uno\ has allowed us to detect extended outflowing ($\rm v_{max}\sim 1300$ \kms) [OIII] emission in the North direction up to a distance of $\sim$ 9 kpc (see Fig. \ref{fig:vel-flussi1326}).
Finally, the low S/N of the "off-nuclear" spectra extracted for J0900$+$4215 and 1549$+$1245 hampers to infer any firm conclusion on the presence of extended emission around these quasars.


\subsection{H$\beta$-based SMBH masses and Eddington ratios}\label{sec:smbh}

The virial mass of the central SMBH (M$_{\rm BH}^{H\beta}$) for our WISSH quasars has been derived from the FWHM of the BLR H$\beta$. The latter has been demonstrated to be a more reliable single-epoch SMBH mass estimator than CIV FWHM \citep{BaskinLaor2005,ShenLiu2012}, which has been adopted for the calculation of the M$_{\rm BH}$ of WISSH quasars in \cite{WeedmanSargsyanLebouteillerEtAl2012}.
More specifically, we have used the FWHM$_{\rm H\beta}^{\rm BLR}$ values derived from the best-fits (see Table \ref{tab:comp}), the intrinsic  luminosity at 5100 \AA,  $\rm \lambda L_{\lambda}(5100$ \AA), obtained from broad band (mid-IR to UV) SED fitting (Duras et al. in preparation, see Table \ref{tab:mbh}), and  the empirical BLR size--luminosity  relation reported in \cite{BongiornoMaiolinoBrusaEtAl2014}, i.e.:
\begingroup\makeatletter\def\f@size{9}\check@mathfonts
\begin{equation}
\rm Log(M_{BH}^{H\beta}/M_\odot)=6.7+2\rm Log\left(\frac{FWHM_{H\beta}}{10^3\rm km\ s^{-1}}\right)+0.5Log\left(\frac{\lambda L_{\lambda}(5100\textnormal{\AA})}{10^{44}\rm erg\ s^{-1}}\right)\hspace{0.2cm}
\end{equation}
\endgroup
The resulting M$_{\rm BH}^{H\beta}$ values\footnote{The error associated to the M$_{\rm BH}^{H\beta}$ includes both the statistical uncertainties affecting the $\rm \lambda L_{\lambda}(5100$ \AA) and FWHM$_{\rm H\beta}^{\rm BLR}$ values and the systematic uncertainty in the virial relation itself \citep[$\sim0.3$ dex, see][for a complete discussion]{BongiornoMaiolinoBrusaEtAl2014}} for the WISSH quasars, ranging from $\sim2\times10^9$ M$_\odot$ up to $\sim1.6\times10^{10}$ M$_\odot$, are shown in Table \ref{tab:mbh} and indicate that our targets populate the massive end of the black hole mass function at $z\sim2.5-3.5$ \citep{KellyMerloni2012}. Table \ref{tab:mbh} also lists the SED-based bolometric luminosity and Eddington ratio $\lambda_{\rm Edd}=\rm L_{Bol}/L_{Edd}$. The WISSH quasars are accreting at a high rate, with $\rm \lambda_{Edd}\sim0.4-3$. This further demonstrates that WISSH quasars clearly offer the opportunity of collecting high-mass, highly accreting SMBHs in a redshift interval that corresponds to the peak of the quasar number density.

\begin{table}[]
	\centering
	\caption{Luminosity at 5100 \AA, bolometric luminosity, SMBH mass and Eddington ratio of the five WISSH quasars. Columns give the following information: (1) SDSS ID, (2) intrinsic luminosity at 5100 \AA\ (in units of $10^{47}$erg s$^{-1}$), (3) bolometric luminosity (in units of $10^{47}$erg s$^{-1}$), (4) H$\beta$-based SMBH mass (in units of 10$^9$ M$_{\odot}$), (5) CIV-based SMBH mass (in units of 10$^9$ M$_{\odot}$) listed in \cite{WeedmanSargsyanLebouteillerEtAl2012} and (6) Eddington ratio, as derived from our  H$\beta$-based SMBH mass estimates.}
	\small
	\begin{tabular}{lccccc}
		\hline
		SDSS       & $\rm \lambda L_{5100\AA}$ & $\rm L_{Bol}$ & M$\rm _{BH}^{H\beta}$ & $\rm M_{BH}^{CIV}$ & $\rm \lambda_{Edd}$ \\
		(1)        & (2)               & (3)           & (4) & (5) & (6)\\
		\hline
		J0745+4734 & 1.8               & 10.0          & 15.7$\pm$2.1   &  8.7$\pm0.5$ & 0.5$\pm$0.1 		   \\
		J0900+4215 & 1.7               & 8.3           & 2.1$\pm$0.3    &  6.6$\pm0.4$ & 3.1$\pm$0.5 		   \\
		J1201+1206 & 1.2               & 5.9           & 6.5$\pm$0.6    &  6.2$\pm0.2$ & 0.7$\pm$0.1         \\
		J1326-0005 & 1.0               & 5.6           & 2.1$\pm$0.3    &  0.11 $\pm0.02$ & 2.1$^{+0.3}_{-0.6}$ \\
		J1549+1245 & 1.3               & 6.5           & 12.6$\pm$2.2   &  1.9$\pm 0.5$ & 0.4$\pm$0.1 \\		
		\hline
	\end{tabular}
	
	\label{tab:mbh}
\end{table}

\begin{figure}[]
	\centering
	\caption{Bolometric luminosity as a function of the H$\beta$-based SMBH mass (M$_{\rm BH}^{H\beta}$) for different AGN samples. The WISSH quasars are indicated by red stars, while squares refer to X-ray selected Type 1 AGN from \cite{LussoComastriSimmonsEtAl2012}. Circles and crosses indicate the optically selected $z\sim2-3$ quasars from \cite{ShemmerNetzerMaiolinoEtAl2004} and \cite{Shen2016}, respectively. The lines correspond to $\rm L_{Bol}/L_{ Edd}=1$ (dashed), $\rm L_{Bol}/L_{Edd}=0.5$ (solid) and $\rm L_{Bol}/L_{Edd}=0.1$ (dotted).} 
	\label{fig:lbolmbh}
	\includegraphics[width=1.0\columnwidth]{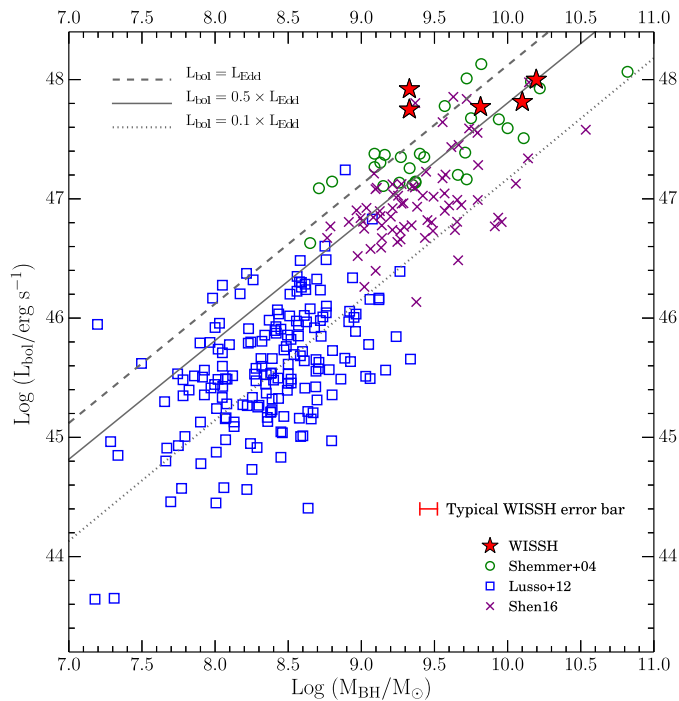}
	
\end{figure}

Apart from \tre, the H$\beta$-based $\rm M_{BH}^{H\beta}$ are significantly  $(\gtrsim3\sigma)$ different from the $\rm M_{BH}$ values derived from the CIV line in \cite{WeedmanSargsyanLebouteillerEtAl2012}, also reported in Table \ref{tab:mbh}. In case of \uno\ and \due, the big difference (larger than a factor of 6 up to 20) between the two estimates can be explained in terms of extinction (which is more effective for UV lines, see Sect. \ref{sec:sed}) or presence of a broad absorption line \citep{GibsonJiangBrandtEtAl2009} which strongly affects the CIV line profile measured in the SDSS spectrum.

Fig. \ref{fig:lbolmbh} shows the bolometric luminosity as a function of  M$\rm _{BH}^{H\beta}$  for the five WISSH quasars, compared with other AGN samples from the literature.
Specifically, we used the sample of $\sim$170  X-ray selected  broad-line AGN in COSMOS \citep[for which the M$\rm _{BH}$ has been derived from H$\beta$ and MgII;][]{LussoComastriSimmonsEtAl2012}  and $\sim$ 100 optically-bright, SDSS quasars with  Log(\lbol/erg s$^{-1}$) $\sim46-48$ at $1.5< z <3.5$, obtained by merging the samples of \cite{Shen2016} and \cite{ShemmerNetzerMaiolinoEtAl2004},
which report H$\beta$-based SMBH masses.
The location of WISSH quasars in the \lbol$-$M$_{\rm BH}^{H\beta}$ plane,  indicates that the WISE selection criterion provides a simple and valuable tool to complete the census of the extreme SMBH population in the universe.


\begin{figure*}[t]
	\centering
	\caption{FWHM$^{\rm broad}_{\rm [OIII]}$ as a function of the total observed [OIII]$\lambda5007$ \AA\ luminosity ($\rm L_{[OIII]}^{Tot}$) for different AGN samples. WISSH quasars (red stars) are compared with other samples of Type 1 and Type 2 AGN at low ($z<1$, empty symbols) and high ($z>1$, filled symbols) redshift. We plot the values obtained for: a sample of obscured \citep{GreeneZakamskaLiuEtAl2009,BrusaBongiornoCresciEtAl2015,PernaBrusaCresciEtAl2015}, red \citep{UrrutiaLacySpoonEtAl2012}, Type 2 \citep{Villar-MartinHumphreyDelgadoEtAl2011,LiuZakamskaGreeneEtAl2013}, luminous $z\sim2.4$ \citep{Cano-DiazMaiolinoMarconiEtAl2012, CarnianiMarconiMaiolinoEtAl2015}, KASHz at $z\sim1.1-1.7$ \citep{HarrisonAlexanderMullaneyEtAl2016} and radio quasars \citep{KimHoLonsdaleEtAl2013}, in addition to the local and high-z ULIRGs by \cite{RodriguezZaurinTadhunterEtAl2013} and \cite{HarrisonAlexanderSwinbankEtAl2012}, respectively. The average values in two $\rm L_{[OIII]}^{Tot}$ bins for optically-selected Type 2 AGN in \cite{MullaneyAlexanderFineEtAl2013} are also shown.}
	\includegraphics[width=0.9\linewidth]{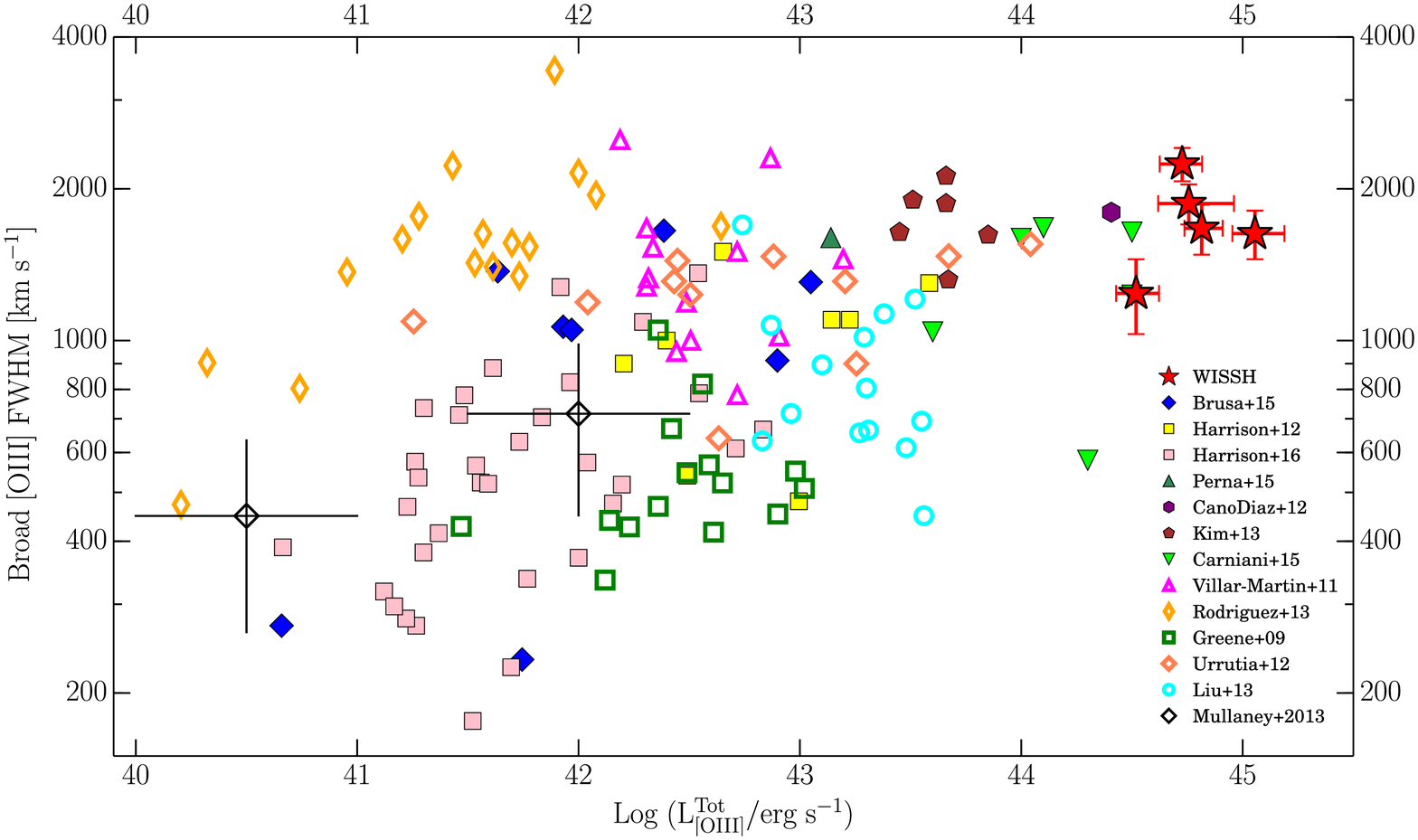}
	
	\label{fig:LOIII}
	
\end{figure*}

\begin{table}[t]
	\centering
	\caption{Properties of the core and broad components of the [OIII] emission, as derived from the best-fits of the LUCI1 spectra (see Sect. \ref{tab:comp}).  Columns give the following information: (1) SDSS ID,  (2) observed [OIII] flux (in units of $10^{-15}$erg cm$^{-2}$s$^{-1}$), (2) observed and (3) intrinsic luminosity  (in units of $10^{44}$erg s$^{-1}$). The latter has been calculated  using $\rm E(B-V)$ values derived in Sect. \ref{sec:sed}.}
	\footnotesize
	\noindent\makebox[1\columnwidth]{
		
		\begin{tabular}{lccc}
			
			\hline
			
			SDSS            &	F$\rm ^{core}_{[OIII]}$     &	L$\rm ^{core}_{[OIII],obs}$	  & L$\rm ^{core}_{[OIII],int}$    \\
			                &   F$\rm ^{broad}_{[OIII]}$    &	L$\rm ^{broad}_{[OIII],obs}$  & L$\rm ^{broad}_{[OIII],int}$   \\
			(1)             &   (2)         	            &	(3)                           & (4) 	                       \\
			\hline
			J0745$+$4734    & 1.8$\pm0.9$                   &   1.7$\pm0.8$                   & 1.7$\pm0.8$      	  \\			
		                 	& 11.9$\pm3.0$               	&   11.4$\pm2.9$                  & 11.4$\pm2.9$  	      \\
			                &                               &                                 &                       \\
			J0900$+$4215    & 0.4$_{-0.2}^{+1.0}$           &   0.5$_{-0.3}^{+1.0}$           & 0.5$_{-0.3}^{+1.0}$   \\			
			                & 5.2$\pm1.0$                   & 	5.3$\pm1.0$                   & 5.3$\pm1.0$  	      \\
			                &                               &                                 &                       \\
			J1201$+$1206    & $(4.3\pm0.9$                  &   5.0$\pm1.0$                   & 5.0$\pm1.0)^a$        \\			
			                & 1.3$\pm0.7$                   &   1.5$\pm0.8$                   & 1.5$\pm0.8$           \\
			                & (1.8$_{-0.4}^{+1.0}$          &   2.1$_{-0.5}^{+1.2}$           & 2.2$_{-0.6}^{+1.3}$)$^b$\\
			                &                               &                                 &                       \\
			J1326$-$0005    &   2.9$\pm1.7$                 &   3.0$\pm1.7$                   & 6.0$\pm$3.1	     	  \\		
			                &   5.6$\pm1.3$                 &   5.7$\pm1.3$                   & 11.4$\pm2.6$          \\
			                &                               &                                 &                       \\
			J1549$+$1245    &   0.3$_{-0.1}^{+0.6}$         &	0.15$_{-0.08}^{+0.30}$        & 0.2$_{-0.1}^{+0.4}$   \\
			                &   6.8$\pm1.2$                 &	3.1$\pm0.7$                   & 4.7$\pm0.9$  	      \\			
			\hline
			
		\end{tabular}
	}
	\flushleft{$^a$ Values derived for the broad [OIII] component at 5007 \AA\ (see Sect. \ref{sec:best}). \\ $^b$ Values derived for the additional broad, shifted [OIII] component included in models C1 and C2 (see Sect. \ref{sec:spectral}).}
	\label{tab:OIIIpar}
\end{table}


\vspace{0.75cm}

\section{Discussion}\label{sec:discuss}

\subsection{Properties of the broad [OIII] emission}

All the five analysed WISSH quasars spectra  show very broad [OIII]  emission ($\rm FWHM_{\rm [OIII]}^{\rm broad}\sim1200-2200$ km s$^{-1}$, see Table \ref{tab:comp}). A systemic, core ($\rm FWHM_{\rm [OIII]}^{\rm core}\lesssim500-700$ km s$^{-1}$) [OIII] component, typically associated with emission from the NLR, is detected in all targets but \tre. However, the  [OIII]$_{\rm broad}$/[OIII]$_{\rm core}$ flux ratios ranging from 2 to 20 indicate that the bulk of the emission is provided by the broad component. In the case of \tre, the emission line centered at 5007 \AA\ exhibits a very large FWHM value ($\sim$ 1700 km s$^{-1}$) and, therefore, can be also considered as representative of outflow.

The broad, shifted [OIII] component is blueshifted in all cases except  \due, for which a mildly redshifted ($\lambda_{\rm [OIII]}^{\rm broad}\sim5012$ \AA) broad component has been found.
Fluxes and luminosities of the core and broad [OIII] emissions derived from the best-fits of the LUCI1 spectra are shown in Table \ref{tab:OIIIpar}. In case of \tre\ we also list the values associated to the additional broad [OIII] component present in model C1 (see Sect. \ref{sec:spectral}).
The values of the intrinsic [OIII] luminosity  have been computed using the SMC extinction curve and the $\rm E(B-V)$ values listed in Table \ref{tab:mag} (see Sect. \ref{sec:sed}).

A comparison with previous studies on ionised outflows reveals that the WISSH quasars survey gives the opportunity to expand the explorable range of [OIII] luminosity up to $\sim10^{45}$ \ergs.
This is shown in Fig. \ref{fig:LOIII}, in which the FWHM of the broad [OIII] line is plotted versus the total observed [OIII]$\lambda5007$ \AA\ luminosity ($\rm L_{\rm [OIII]}^{\rm Tot}$). WISSH quasars are compared with various AGN samples, including both Type 1 and Type 2 objects at low ($ z<$ 1) and high ($ z>$ 1) redshift.
The WISSH quasars are located at the right-top corner of the plot, indicating that our selection allows to extend the discovery of ionised outflows up to sources with observed $\rm L_{[OIII]}^{\rm Tot}\gtrsim3\times10^{44}$ \ergs up to $\sim10^{45}$ \ergs in case of \quattro.
There is a hint that the FWHM$_{\rm [OIII]}^{\rm broad}$ increases with the [OIII] luminosity \citep{MullaneyAlexanderFineEtAl2013}, while no redshift dependence seems to be present. 
More specifically, a lack of AGN  with  luminous [OIII] emission exhibiting a narrow ($<$ 1000 \kms) line profile is evident in the $\rm FWHM_{\rm [OIII]}^{\rm broad}-\rm L_{\rm[OIII]}^{\rm Tot}$ plane.
This finding is further supported by the fact that in all other thirteen WISSH quasars observed with LUCI1,  the [OIII] emission is absent or  very weak, with $\rm L_{[OIII]}^{\rm Tot}\lesssim 10^{43}$ \ergs\ (details about these observations will be discussed in a forthcoming paper by Vietri et al.).


\subsection{Mass Rate and Kinetic Power of Ionised Outflows}\label{sec:mdot}

In the following, we calculate the mass rate ($\rm \dot{M}$) and kinetic power ($\rm \dot{E}_{kin}$) of the ionised outflows in our WISSH quasars, using the [OIII] emission line as ionised gas mass tracer, similarly to many previous studies \citep[e.g.][]{HarrisonAlexanderSwinbankEtAl2012,HarrisonAlexanderMullaneyEtAl2014,Cano-DiazMaiolinoMarconiEtAl2012,LiuZakamskaGreeneEtAl2013,BrusaBongiornoCresciEtAl2015,KakkadMainieriPadovaniEtAl2016}.
However, we are aware that [OIII] is not a robust mass tracer, as it is very sensitive to gas temperature (and, therefore, metallicity), ionisation parameter and electron density.
Nonetheless, for our targets this emission line is the only observed transition not affected by BLR motions and, therefore, the only available tracer to estimate (at least roughly) the ionised outflowing gas properties.

We assume a spherically/biconically symmetric mass-conserving free wind, with mass outflow rate and velocity that are independent of radius as in \cite{RupkeVeilleux2002,RupkeVeilleuxSanders2005}.
We derive the outflowing ionised gas mass from the luminosity associated to the broad [OIII] emission, using the following relation from \cite{CarnianiMarconiMaiolinoEtAl2015} under the strong assumption that most of the oxygen consists of $\rm O^{2+}$ ions:
\begin{equation}\label{eq:mion}
\rm M_{ion}^{out}=4.0\times 10^7M_\odot\left(\frac{C}{10^{[O/H]}}\right)\left(\frac{L_{[OIII]}}{10^{44}\ erg\ s^{-1}}\right)\left(\frac{\langle n_e\rangle}{10^3\ cm^{-3}}\right)^{-1}
\end{equation}
\noindent where $\rm C=\langle n_e\rangle^2/\langle n_e^2\rangle$ and $\rm n_e$ is the electron density.
\begin{table*}[]
	\centering
	\caption{Properties of the outflows discovered in the five WISSH quasars. Columns give the following information: (1) SDSS ID, (2) velocity shift (in units of \kms), (3) maximum outflow velocity (in units of \kms), (4) ionised outflowing gas mass (in units of $10^{9}$ M$_{\odot}$), (5) mass outflow rate (in units of M$_{\odot}$yr$^{-1}$), (6) kinetic power (in units of $10^{45}$ erg s$^{-1}$) of the outflow, (7) ratio between the kinetic power of the outflow and the bolometric luminosity of the quasar. See Sect. \ref{sec:mdot} for a detailed description of the assumptions made to derive M$_{\rm ion}^{\rm out}$, $\rm \dot{M}$ and $\rm \dot{E}_{kin}$.}
	\footnotesize
	\begin{tabular}{lccccccc}
		
		\hline
		SDSS         &     & $\Delta$v$^a$	 & v$^{b}_{\rm max}$ & M$_{\rm ion}^{\rm out}$  & $\rm \dot{M}$  & $\rm \dot{E}_{kin}$  & $\rm \dot{E}_{kin}/L_{Bol}$  \\
		(1) & & (2) & (3) & (4) & (5) & (6)  & (7)   \\
		\hline
		J0745$+$4734 &     & $-$510$\pm$150  & 1890$\pm$170      & 6.9            & 5700           & 6.7  				  & 0.007  \\
		J0900$+$4215 &     & $-$480$\pm$160  & 2380$\pm$180      & 3.2            & 3300           & 6.3 			      & 0.008  \\
		J1201$+$1206 &(I)$^c$  & $-$	         & 1420$\pm$160      & 3.1            & 2500           & 2.8  			      & 0.005  \\
		&(II)$^c$ & $-$1060$\pm$120 & 1850$\pm$180      & 0.9            & 740            & 0.9  			      & 0.001  \\
		&(III)$^c$ & $-$1990$\pm$200 & 3270$\pm$230      & 1.3            & 1890           & 6.7  				  & 0.012  \\
		J1326$-$0005 &     & $-$580$\pm$100  & 2160$\pm$170      & 8.2            & 7740           & 12.0 				  & 0.031  \\
		J1549$+$1245 &     & $+$330$\pm$130  & 1380$\pm$220	     & 2.9            & 1740           & 1.1 				  & 0.002  \\
		\hline
	\end{tabular}
	
	\flushleft
	$^a$ The velocity shift $\Delta$v of the broad [OIII] component was derived from the shift in wavelength ($-$ sign indicates a blueshift, $+$ sign indicates a redshift)  between broad/shifted and narrow/systemic [OIII] centroids.\\
	$^b$ The outflow maximum velocity has been computed as $\rm v_{max}=|\Delta v|+2\sigma_{[OIII]}^{broad}$.\\
	$^c$ Values derived (I) for the broad component centred at 5007 \AA\ (see Sect. \ref{sec:best}); (II) for the broad blueshifted component centred at $\lambda\sim4990$ \AA; (III) for the additional broad blueshifted component centred at $\lambda\sim4970$ \AA.
	\label{table:outpar}
\end{table*}
According to this simple model, in which an ionised cloud uniformly filled up to a certain radius R, is ejected outward of the central regions in the quasar host galaxy, and  considering that the mass outflow rate is equal to the total ionised gas mass divided the dynamical timescale (the time necessary for the mass $\rm M_{ion}^{out}$ to pass through the sphere of radius R), we obtain:
\begin{equation}
\rm \dot{M}\approx\frac{M_{ion}^{out}v}{R}\hspace{0.2cm}
\end{equation}
\noindent where $\rm v$ is the outflow velocity. Therefore:
\begingroup\makeatletter\def\f@size{7.7}\check@mathfonts

\begin{equation}
\rm \dot{M}\approx41M_\odot yr^{-1} \left( \frac{C}{10^{[O/H]}} \right) \left( \frac{L_{[OIII]}}{10^{44}erg\ s^{-1}}\right) \left(\frac{\langle n_e\rangle}{10^3cm^{-3}}\right)^{-1} \left(\frac{v}{10^3 km\ s^{-1}}\right) \left( \frac{R}{1 kpc}\right)^{-1}
\label{eq:mdot-tot}
\end{equation}
\endgroup 
Another estimate of $\rm \dot{M}$ is provided by the fluid field continuity equation: if $\rm \rho=\frac{3M_{ion}^{out}}{\Omega\pi R^3}$ is the mean density of an outflow covering the solid angle $\rm \Omega\pi$, then the $\rm \dot{M}$ is given by:
\begin{equation}
\rm \dot{M}=\Omega\pi R^2\rho v=3\frac{M_{ion}^{out}v}{R}\hspace{0.2cm},
\label{eq:mdot-loc}
\end{equation}
\noindent i.e. a factor of 3 larger than the value derived from Eq. \ref{eq:mdot-tot}. This is a local estimate of $\rm \dot{M}$ at a given radius R \citep[e.g.][]{FeruglioFioreCarnianiEtAl2015}.
From $\rm \dot{M}$, the kinetic power associated to the outflow can be derived as follows:
\begin{equation}\label{eq:edotkin}
\rm \dot{E}_{kin}=\frac{1}{2}\dot{M}v^2\hspace{0.2cm}.
\end{equation}

\begin{figure}[]
	\centering	
	\caption{Maximum velocity of the outflow versus the AGN \lbol\ of the WISSH quasars compared with other samples from literature \citep[i.e.][]{HarrisonAlexanderSwinbankEtAl2012,HarrisonAlexanderMullaneyEtAl2014,GenzelFoersterSchreiberRosarioEtAl2014,CarnianiMarconiMaiolinoEtAl2015,PernaBrusaCresciEtAl2015,BrusaPernaCresciEtAl2016,NesvadbaLehnertEisenhauerEtAl2006,NesvadbaLehnertDeBreuckEtAl2008,RupkeVeilleux2013}. The dashed line represents the relation $\rm L_{Bol}\propto v_{max}^5$ (with arbitrary normalization) expected for energy conserving winds \citep[see][]{CostaSijackiHaehnelt2014}.}
	\includegraphics[width=1\columnwidth]{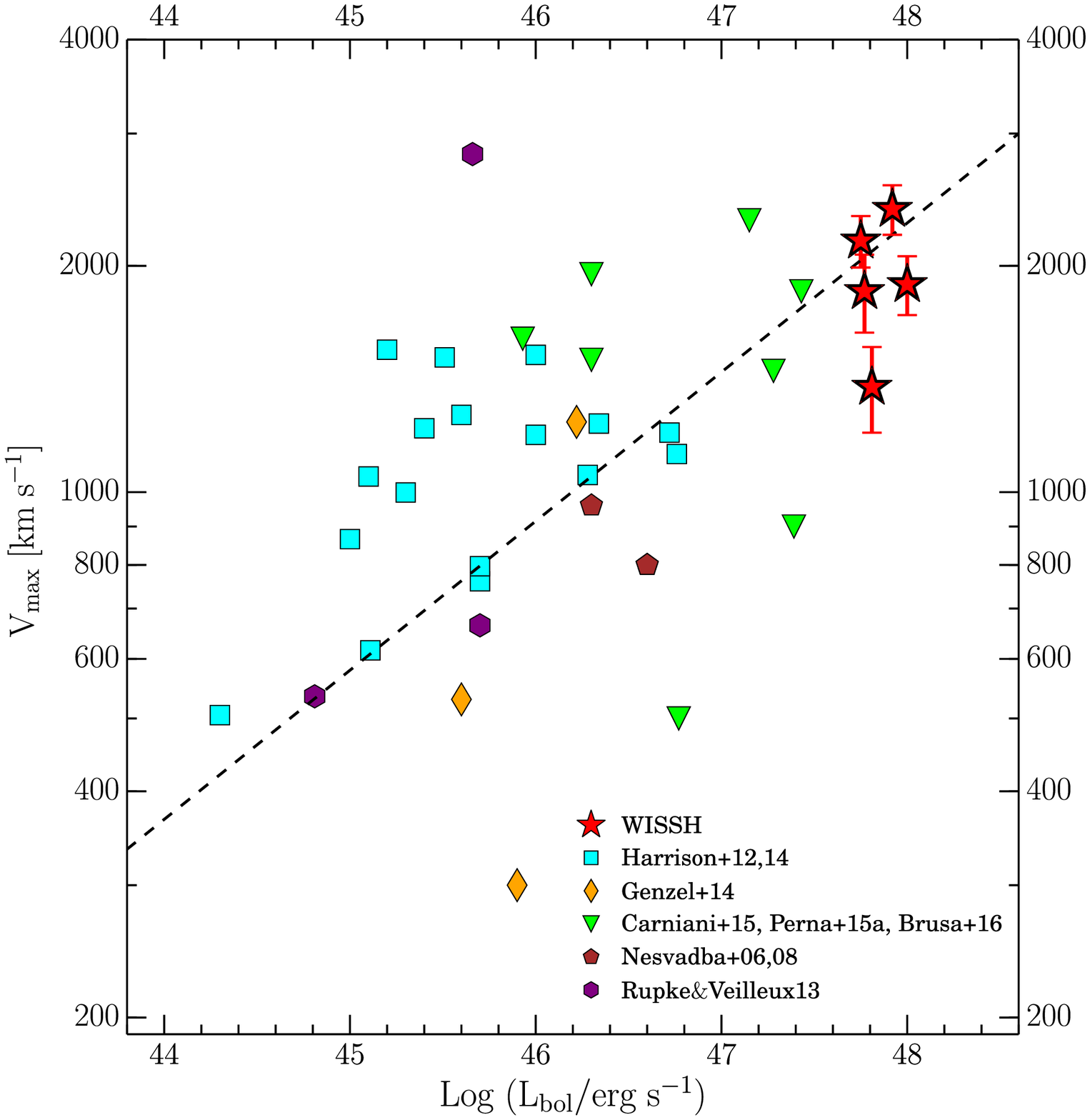}
	\label{fig:vmax-lbol}
\end{figure}

\begin{figure*}[]
	\centering
	\caption{Mass rate \textit{(a)} and kinetic power \textit{(b)} of the ionised outflows as a function of \lbol. Values obtained for the WISSH quasars (red stars) are compared with other samples from literature \citep[i.e.][]{HarrisonAlexanderSwinbankEtAl2012,HarrisonAlexanderMullaneyEtAl2014,GenzelFoersterSchreiberRosarioEtAl2014,CarnianiMarconiMaiolinoEtAl2015,PernaBrusaCresciEtAl2015,BrusaPernaCresciEtAl2016,NesvadbaLehnertEisenhauerEtAl2006,NesvadbaLehnertDeBreuckEtAl2008,RupkeVeilleux2013}. Error bars are calculated as described in Sect. \ref{sec:mdot}. In panel \textit{(a)} the dotted line corresponds to the best-fit relation $\rm \dot{M}\propto L_{Bol}^{1.21}$ derived for ionised outflows and the dashed line to the $\rm \dot{M}\propto L_{Bol}^{0.78}$ best-fit relation to molecular outflows \citep[see][for a complete discussion]{FioreEtAl2016}. In panel \textit{(b)} dashed, dotted and solid lines represent an outflow kinetic power that is 10\%, 1\% and 0.1\% of the AGN luminosity, respectively.}
	\makebox[1\textwidth]{
		\subfigure[]{
			\includegraphics[width=0.5\textwidth]{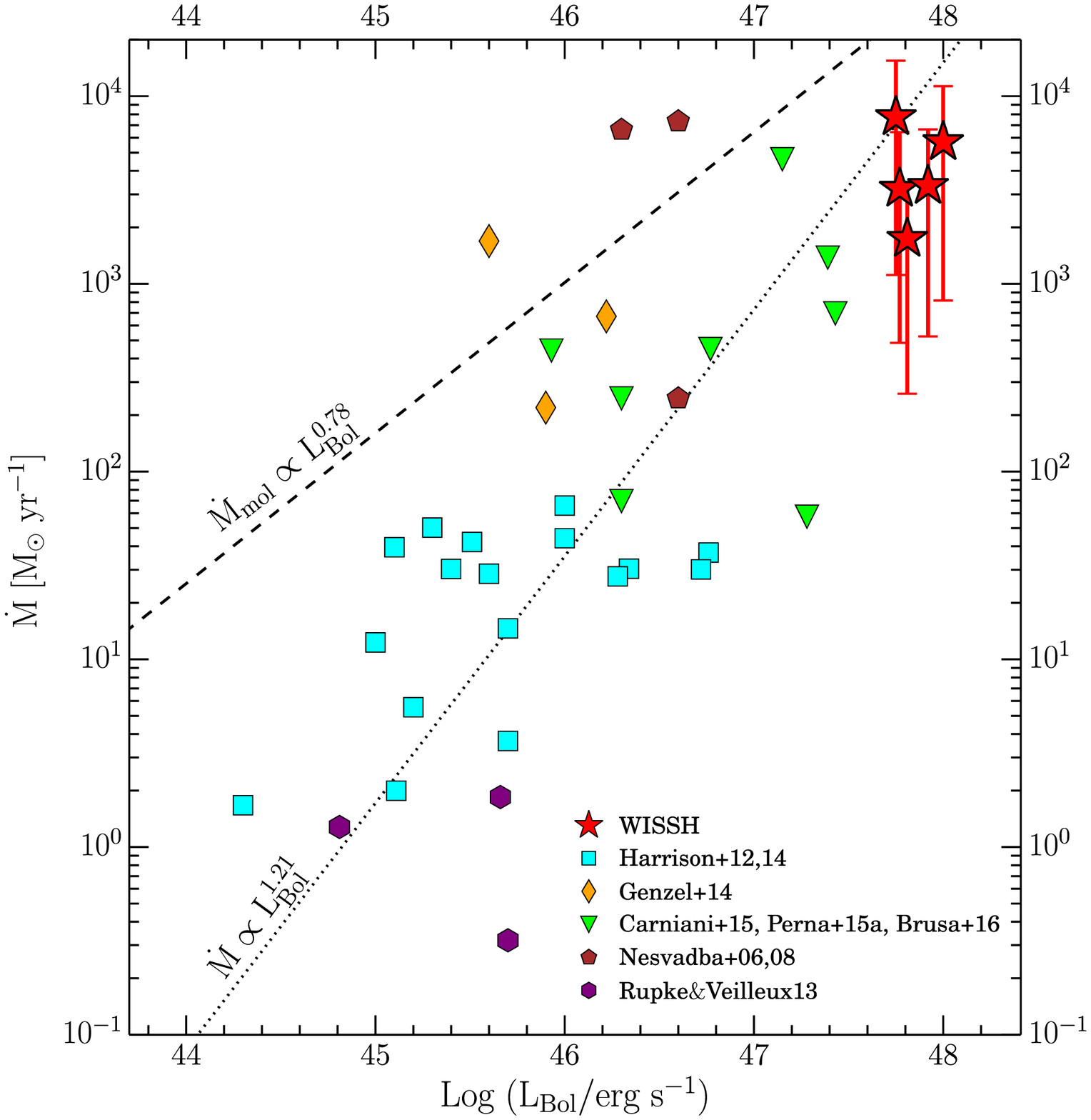}}
		\subfigure[]{
			\includegraphics[width=0.5\textwidth]{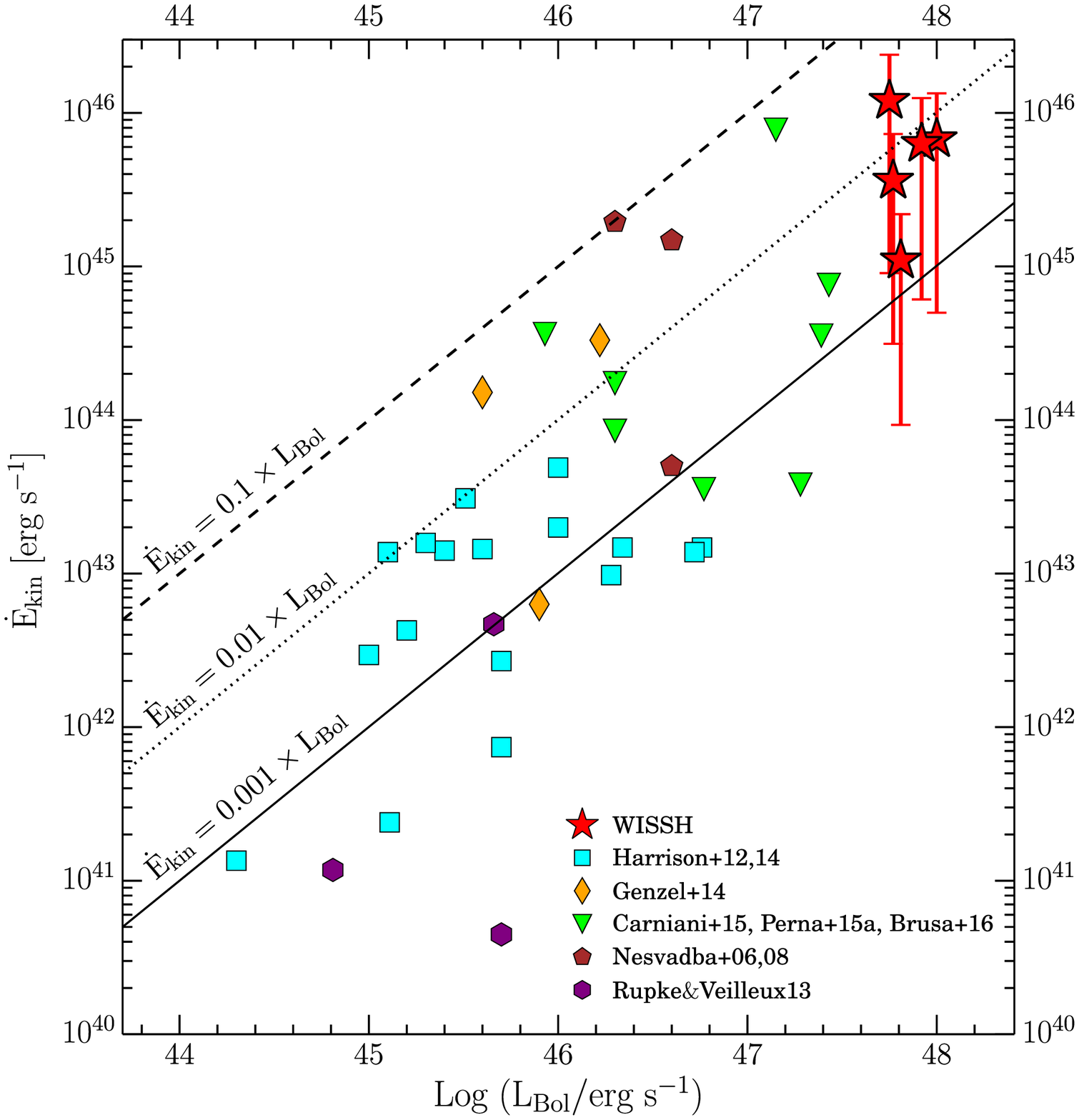}}
	}
	\label{fig:mdotlboli}
\end{figure*}

\noindent  We compute the $\rm \dot{M}$ and $\rm \dot{E}_{kin}$ for our WISSH quasars according to Eq. \ref{eq:mdot-loc} and Eq. \ref{eq:edotkin}, respectively. More specifically, we assume $\rm C\approx1$, $\rm [O/H]\sim0$ (solar metallicity) and an electron density $\rm n_e=200$ cm$^{-3}$, as it represents the average value among the samples presented in Fig. \ref{fig:vmax-lbol} and Fig. \ref{fig:mdotlboli}. Typical [SII]-based measures of electron density vary from $\rm n_e\sim100$ \citep[e.g.][]{PernaBrusaCresciEtAl2015} up to $\rm n_e\sim500$ cm$^{-3}$ \citep{NesvadbaLehnertEisenhauerEtAl2006}. According to our simple model, the maximum velocity $\rm v_{max}=|\Delta v| + 2\sigma_{[OIII]}^{broad}$  (where $\rm \Delta v$ is the velocity shift  between broad and core [OIII] emission centroids), can be considered as representative of the bulk velocity of the outflowing gas. The $\rm v_{max}$ represents the outflow velocity (assumed constant with radius and spherically symmetric) along the line of sight. In our calculations of outflow parameters, we use $\rm v_{max}$ \citep[e.g.][]{GenzelNewmanJonesEtAl2011,Cano-DiazMaiolinoMarconiEtAl2012,BrusaBongiornoCresciEtAl2015} and the intrinsic broad [OIII] luminosity values listed in Table \ref{table:outpar} and Table \ref{tab:OIIIpar}, respectively.

We note that values provided here represent a rough estimate of the $\rm \dot{M}$ and $\rm \dot{E}_{kin}$ of the ionised wind  since Eq.  \ref{eq:mion} is based on several assumptions which lead to an uncertainty level on $\rm \dot{M}$ up to an order of magnitude. A comprehensive description of these assumptions and their impact on the determination of the outflow parameters is presented in \cite{KakkadMainieriPadovaniEtAl2016}, to which we refer the reader for details.

Furthermore, because of the lack of accurate information on the spatial scale of the outflowing gas, we assume R $\sim$ 7 kpc as the radius of the emitting broad [OIII] region. We consider this a reasonable assumption based on the results presented in Sect. \ref{sec:2d}, as it corresponds to the spatial scale at which we observe fast extended [OIII] in \tre\ and \uno. We stress that this choice implies conservative values of $\rm \dot{M}$ and $\rm \dot{E}_{kin}$, as it maximizes the spatial scale of the outflows. Smaller values of the radius would lead to even higher values (see Eq. \ref{eq:mdot-loc}). 
\vspace{0.5cm}

Fig. \ref{fig:vmax-lbol}, Fig. \ref{fig:mdotlboli}a and Fig. \ref{fig:mdotlboli}b show the values derived for our WISSH quasars of $\rm v_{max}$, $\rm \dot{M}$ and $\rm \dot{E}_{kin}$ as a function of \lbol, respectively\footnote{In case of \tre, we plot the $\rm v_{max}$, $\rm \dot{M}$ and $\rm \dot{E}_{kin}$ values obtained from the best fit without considering the additional broad, blue-shifted OIII component included in model C1 (see Sect. \ref{sec:analysis}). Including the additional component would lead to larger $\rm \dot{M}$ and $\rm \dot{E}_{kin}$ (see Table \ref{table:outpar}).}.
We compare our findings with those of: (i) obscured X-ray selected quasars studied by \cite{BrusaPernaCresciEtAl2016} and \cite{PernaBrusaCresciEtAl2015}; (ii) [OIII]-loud, hyper-luminous ($\rm L_{Bol}>10^{47}$ erg s$^{-1}$) quasars at $z\sim1.5-2.5$ in \cite{CarnianiMarconiMaiolinoEtAl2015};  (iii) high-z radio-galaxies analysed by \cite{NesvadbaLehnertEisenhauerEtAl2006,NesvadbaLehnertDeBreuckEtAl2008}; (iv)  low-z and high-z (mostly Type 2) AGN with [OIII] outflows from \cite{HarrisonAlexanderSwinbankEtAl2012,HarrisonAlexanderMullaneyEtAl2014}; (v) massive AGN at $z\sim2$ from \cite{GenzelFoersterSchreiberRosarioEtAl2014}.

Fig. \ref{fig:vmax-lbol} suggests a correlation between maximum outflow velocity and AGN luminosity, with all the ionised outflows discovered in WISSH quasars exhibiting among the largest $\rm v_{max}$ values ($\gtrsim1300$ km s$^{-1}$). As previously reported by \cite{Shen2016},  \cite{ZakamskaGreene2014} and \cite{ZakamskaHamannParisEtAl.2016},  the [OIII] emission lines detected in luminous AGN tend to have larger widths. Interestingly, the observed values are consistent with a  $\rm L_{Bol}\propto v_{max}^5$ relation, which is expected for an energy conserving wind \citep{CostaSijackiHaehnelt2014}.

Some samples included in Fig. \ref{fig:mdotlboli} have published $\rm \dot{M}$ values derived from the luminosity of  H$\alpha$ or H$\beta$ instead of [OIII] luminosity. This leads to values of $\rm \dot{M}$ systematically larger by a factor of $\sim$ 3 up to 10 than those computed from the [OIII] \citep{FioreEtAl2016}. This discrepancy can be due to the fact that Balmer and [OIII] lines are produced in regions with different C and $\rm n_e$ values and to the assumption that $\rm n(O^{2+})/n(O)\sim1$ in the derivation of Eq. \ref{eq:mion} \citep{CarnianiMarconiMaiolinoEtAl2015}.
Accordingly, in order to enable a proper comparison, all the [OIII] based ionised gas masses (and, hence, $\rm \dot{M}$ and $\rm \dot{E}_{kin}$) have been multiplied by a conservative factor of 3 (including those of our WISSH quasars). We note that the alternative estimate of the mass outflow rate from the broad H$\beta$ in our WISSH quasars is a factor of $\sim3.2$ larger than the $\rm \dot{M}$ derived from the [OIII] emission. 
Furthermore, in order to compare  homogeneous values,  all $\rm \dot{M}$ and $\rm \dot{E}_{kin}$ shown in Fig. \ref{fig:mdotlboli}a,b have been inferred using the same assumptions adopted for WISSH quasars, (i.e. $\rm n_e=200$ cm$^{-3}$ and $\rm v_{max}$). Our objects exhibit huge values of $\rm \dot{M}$, i.e. from $\sim 1.7\times10^3$ up to $\sim7.7\times10^3$ \msun\ yr$^{-1}$, and $\rm \dot{E}_{kin}\sim10^{45}-10^{46}$ \ergs (see Table \ref{table:outpar}), which are among the largest ones measured so far.

In order to give an idea of the potential impact of the uncertainties affecting $\rm \dot{M}$ and $\rm \dot{E}_{kin}$, we report error bars for WISSH quasars in Fig. \ref{fig:mdotlboli} estimated as follows: the upper bounds correspond to the assumption of $\rm n_e=80$ cm$^{-3}$ \citep[as in][]{GenzelFoersterSchreiberRosarioEtAl2014}, while the lower bounds correspond to $\rm n_e=1000$ cm$^{-3}$ \citep[i.e. the typical value for the NLR, ][]{Peterson1997} and a velocity of $\rm W_{80}/1.3$\footnote{$\rm W_{80}$ indicates the velocity width of the line at the 80\% of the line flux. For our sources, $\rm W_{80}$ is a factor of 1.3--1.8 smaller than $\rm v_{max}$.}  \citep[as in][]{HarrisonAlexanderMullaneyEtAl2014}. However, we stress that these estimates of the outflow parameters are based assuming a very simplistic scenario, due to the limited information on the physical properties of the [OIII] emitting material.

Both $\rm \dot{M}$ and $\rm \dot{E}_{kin}$ seem to exhibit a correlation with \lbol, although with a large scatter. We note that WISSH quasars show less scatter than other samples in Fig. \ref{fig:mdotlboli}. The increase of mass outflow rates with increasing \lbol\ is consistent with a theoretical scenario according to which the more luminous the AGN, the more is the amount of gas invested by a radiatively-driven wind \citep{MenciFiorePuccettiEtAl2008}. However, a direct relation between $\rm v_{max}$ and \lbol\ is difficult to be established via observations, as the fraction of the outflow kinetic power injected into the ISM may be different from object to object. Moreover, it should be noticed that the \lbol\ derived here may not represent the long time-scale average luminosity of the AGN, responsible for the outflow acceleration \citep{FaucherGiguereQuataert2012}. All these effects can produce the observed large scatter.

Fig. \ref{fig:mdotlboli}a suggests that ionised outflows, revealed at the highest luminosities, may trace a larger fraction of the total outflowing gas than outflows in low luminosity AGN. The best-fit relation $\rm \dot{M}\propto L_{Bol}^{1.21}$ for the ionised outflows is steeper than the relation $\rm \dot{M}_{mol}\propto L_{Bol}^{0.78}$ obtained for molecular outflows \citep[see ][for a complete discussion]{FioreEtAl2016}, indicating that at $\rm L_{bol}\gtrsim10^{47}$ the mass rates of ionised and molecular outflows become comparable. This is also supported by the fact that most of $\rm \dot{E}_{kin}$ values derived for our WISSH quasars are broadly consistent with a fraction of $\sim1$\% of \lbol\ and remarkably, in case of \uno, the ratio $\rm \dot{E}_{kin}/L_{Bol}$ reaches a value of $\sim3$\%. 
These values are therefore very close to the predictions (i.e. $\rm \dot{E}_{kin}/L_{Bol}\sim5$\%) of the vast majority of AGN feedback models \citep[e.g.][]{DiMatteoSpringelHernquist2005,ZubovasKing2012} for an efficient feedback mechanism being able to account for the $\rm M_{BH}$--$\rm \sigma$ relation.
Similar fractions have been indeed observed for molecular outflows \citep{FeruglioMaiolinoPiconcelliEtAl2010,CiconeMaiolinoSturmEtAl2014}, whereas ionised outflows usually show values $< 1$\% of \lbol\ as illustrated in Fig. \ref{fig:mdotlboli}b.

Finally, we note that assuming a density profile $\rm n_e\propto r^{-2}$ (where r is the distance from the galaxy center) as for an isothermal scenario, leads to mass outflow rates reduced by a factor of $\sim 3$ with respect to the values derived considering a uniform density. However, all  $\rm \dot{M}$ and $\rm \dot{E}_{kin}$ plotted in Fig. \ref{fig:mdotlboli} should be reduced by a similar factor, and the positive trend with increasing \lbol\ therefore remains valid.

\section{Summary and Conclusions}\label{sec:conclusions}

In this work, we have analysed  the LBT/LUCI1 optical rest-frame spectra of five  broad-line quasars at $z \sim 2.5-3.5$, with the aim of characterising nuclear properties and AGN-driven outflows at the brightest end of the AGN luminosity function. 
Our targets belong to the WISE/SDSS selected hyper-luminous (WISSH) quasars sample and all have \lbol\  $\gtrsim6\times10^{47}$ \ergs.
We have traced the ionised gas kinematics from the broad wings of the [OIII] emission line and estimated the mass of the SMBH from the BLR H$\rm \beta$ emission. Our key findings can be summarised as follows:

\begin{itemize}

\item Our spectral analysis reveals the presence of very broad (FWHM$\rm _{[OIII]}^{broad}\sim1200-2200$ \kms) [OIII] emission lines (among the largest measured so far), with exceptional luminosities $\rm L_{[OIII]}^{Tot}\gtrsim5\times10^{44}$ \ergs\ (see Fig. \ref{fig:LOIII}). This finding suggests that our WISE/SDSS selection allows to extend the  study of ionised outflows up to $\rm L_{[OIII]}^{broad}\gtrsim10^{45}$ \ergs.
We also find that the WISSH quasars with the largest [OIII] luminosity ($>10^{45}$ erg s$^{-1}$) typically show a small ratio ($\sim2-7$) between the luminosity of the broad and narrow, systemic [OIII] component.

\item The "off-nuclear" spectra of three out of five quasars, extracted at increasing distances from the central source, show hints of extended [OIII] emission on scale of $\sim7-10$ kpc, associated to outflowing gas in \tre\ and \uno\ (see Sect. \ref{sec:2d}).  Furthermore, we are able to  reveal a likely bipolar outflow in \tre, with a  redshifted(blueshifted) [OIII] component seen North(South) off the nucleus out to $\sim$ 7 kpc.

\item Our WISSH quasars host SMBHs with very large H$\beta$-based masses ($\rm M_{BH}^{H\beta}\gtrsim2\times10^9$ M$_{\odot}$), which are accreting at high rates ($\rm \lambda_{Edd}$ from $\sim$ 0.4 up to 3). This suggests  that WISSH quasars  are highly accreting SMBHs at the massive end of the SMBH mass function (see Sect. \ref{sec:smbh}).

\item  We find ionised outflows exhibiting huge values of $\rm \dot{M}\sim1700-7700$ M$_\odot$ yr$^{-1}$ and  $\rm \dot{E}_{kin}\sim10^{45}-10^{46}$ \ergs, indicating that WISSH quasars allow to reveal extreme outflows. We derive $\rm \dot{M}$ values which are closer to the best-fit $\rm \dot{M}_{mol}-L_{Bol}$ relation, derived for galaxy-wide molecular outflows in \cite{FioreEtAl2016}, than  values typically observed in lower AGN luminosity (see Fig. \ref{fig:mdotlboli}). This finding, in addition to the large $\rm \dot{E}_{kin}$/\lbol $\sim$ $0.01-0.03$, suggests that  the ionised outflows in hyper-luminous quasars may trace a substantial fraction of the outflowing gas.
However, it should bear in mind that our estimates of $\rm \dot{M}$ and $\rm \dot{E}_{kin}$ are affected by order-of-magnitude uncertainties, due to the use of [OIII] as tracer of the ionized  outflows and the very basic outflow model we assumed \citep[see sect 6.2 and][]{KakkadMainieriPadovaniEtAl2016}.
\end{itemize}

This paper presents the first results about the presence of powerful ionised winds in WISSH quasars. A complete description of the  optical rest-frame spectral  properties for a larger sample of WISSH quasars covered by LUCI1 observations will be provided in a forthcoming paper (Vietri et al.).
Furthermore, a  follow-up program of \uno\ and \tre\ with the VLT SINFONI integral field spectrograph (P.I. Bongiorno) has been recently accepted.
These observations will  enable us to accurately constrain  the  morphology and, in turn,  the effects on the host galaxy ISM of these  energetic  outflows.

\vspace{0.9cm}

\begin{acknowledgements}
We are grateful to the anonymous referee for valuable feedback that helped to improve the paper.

Based on observations obtained at the LBT. 
The LBT is an international collaboration among institutions in the United States, Italy and Germany. LBT Corporation partners are: The University of Arizona on behalf of the Arizona Board of Regents; Istituto Nazionale di Astrofisica, Italy; LBT Beteiligungsgesellschaft, Germany, representing the Max-Planck Society, the Astrophysical Institute Potsdam, and Heidelberg University; The Ohio State University, and The Research Corporation, on behalf of The University of Notre Dame, University of Minnesota and University of Virginia.
 
This research has made use of the NASA$/$IPAC
Extragalactic Database (NED) which is operated by the Jet Propulsion
Laboratory,  California Institute of Technology, under contract with
the National Aeronautics and Space Administration.
We acknowledge financial support from PRIN-INAF 2014.
A. Bongiorno and E. Piconcelli acknowledge financial support from INAF under the contract PRIN-INAF-2012.
M. Brusa acknowledges support from the FP7 Career Integration Grant "eEasy" (CIG 321913).
R. Schneider and M. Bischetti acknowledge funding from the ERC FP7 (Grant Agreement n. 306476).
L. Zappacosta acknowledges financial support2D under ASI/INAF contract I/037/12/0.
We thank E. Lusso and S. Carniani for providing us with \lbol\ and $\rm M_{\rm BH}$ values for the COSMOS sources, and the $\rm FWMH_{OIII}$ and $\rm L_{OIII}$ values plotted in Fig. \ref{fig:lbolmbh} and  Fig. \ref{fig:LOIII}, respectively.
We also thank S. Gallerani for kindly providing the extinction curve from \cite{GalleraniMaiolinoJuarezEtAl2010} in digital form.
We are grateful to M. Fumana for the assistance in data reduction and M. Perna for useful discussion on off-nuclear spectra.

\end{acknowledgements}


\onecolumn
\section*{Appendix}

\begin{figure*}[]
	\centering
	\caption{Same as Fig. \ref{fig:vel-flussi0745} but for \cinque\ (1 pixel $\sim$ 1.9 kpc at the redshift of this source).}
			\vspace{-0.2cm}
			\subfigure[]{\includegraphics[width=0.69\linewidth]{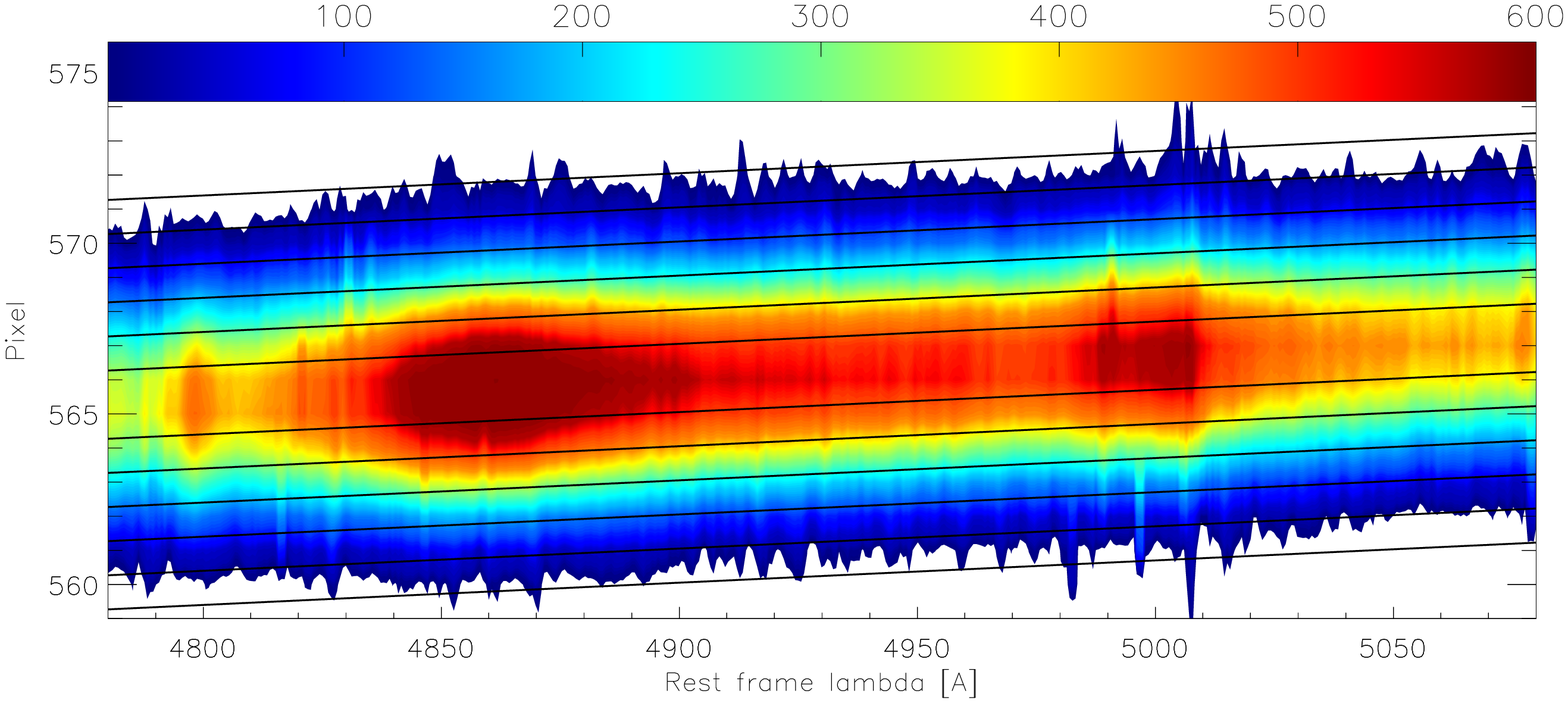}}
			\includegraphics[width=0.95\linewidth]{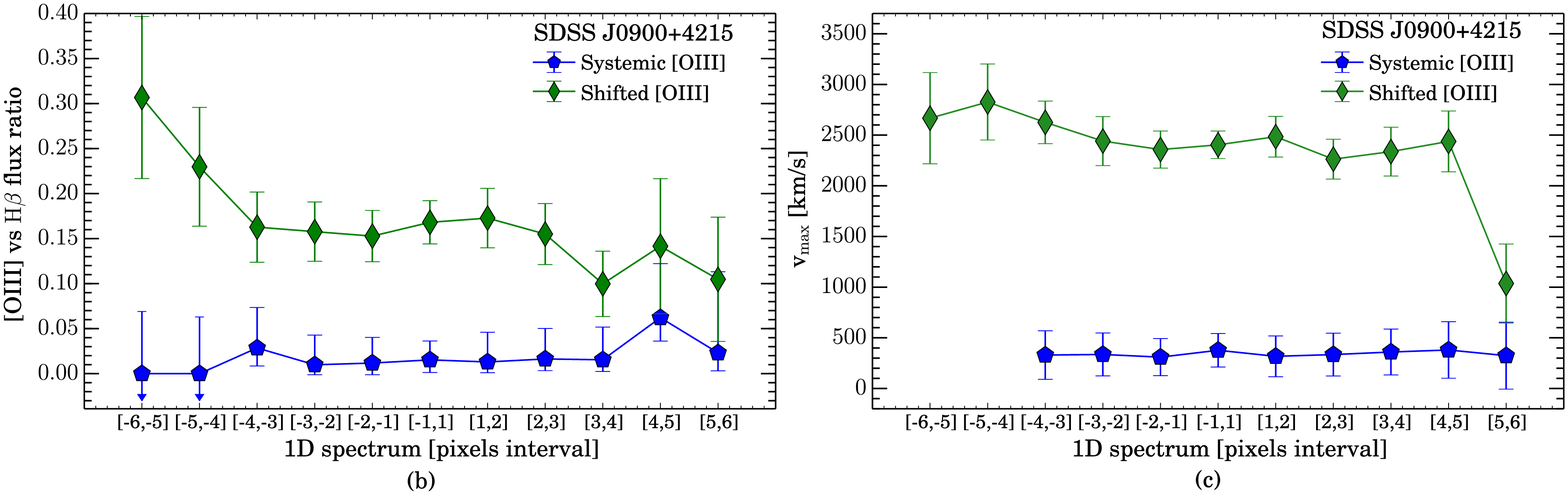}
			\vspace{-0.3cm}
	\label{fig:vel-flussi0900}
\end{figure*}

\begin{figure*}[]
	\centering
	\caption{Same as Fig. \ref{fig:vel-flussi0745} but for \due\ (1 pixel $\sim$ 2.1 kpc at the redshift of this source).}
			\subfigure[]{\includegraphics[width=0.7\linewidth]{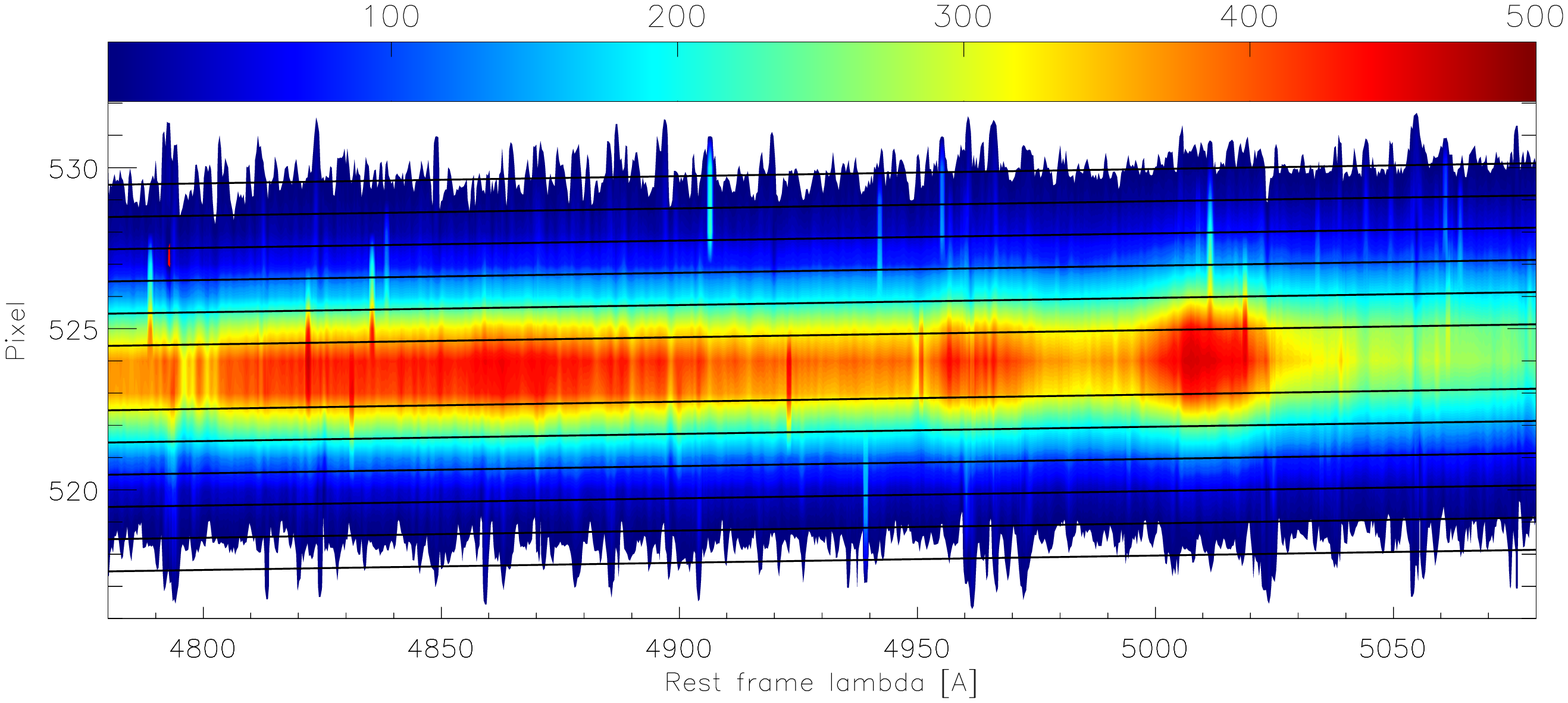}}
			\includegraphics[width=0.95\linewidth]{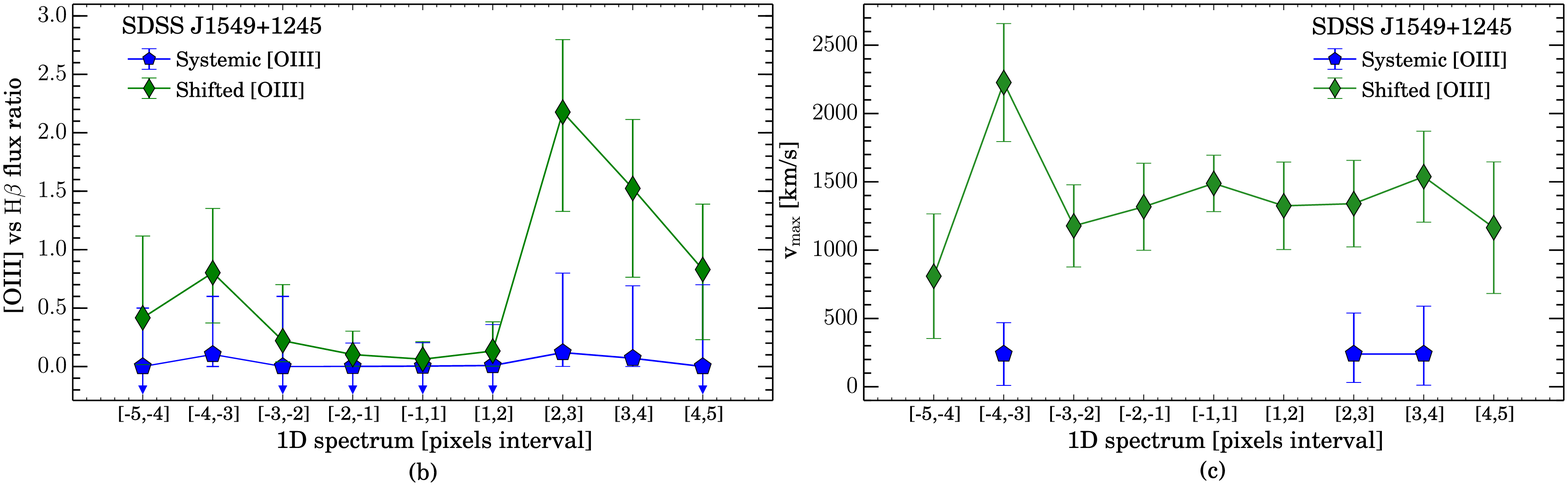}
	\label{fig:vel-flussi1549}
\end{figure*}


\begin{thebibliography}{81}
	\expandafter\ifx\csname natexlab\endcsname\relax\def\natexlab#1{#1}\fi
	
	\bibitem[{{Ahn} {et~al.}(2014){Ahn}, {Alexandroff}, {Allende Prieto}, {Anders},
		{Anderson}, {Anderton}, {Andrews}, {Aubourg}, {Bailey}, {Bastien}, \&
		et~al.}]{AhnAlexandroffAllendeEtAl2014}
	{Ahn}, C.~P., {Alexandroff}, R., {Allende Prieto}, C., {et~al.} 2014, \apjs,
	211, 17
	
	\bibitem[{{Baskin} \& {Laor}(2005)}]{BaskinLaor2005}
	{Baskin}, A. \& {Laor}, A. 2005, \mnras, 356, 1029
	
	\bibitem[{{Bongiorno} {et~al.}(2014){Bongiorno}, {Maiolino}, {Brusa},
		{Marconi}, {Piconcelli}, {Lamastra}, {Cano-D{\'{\i}}az}, {Schulze},
		{Magnelli}, {Vignali}, {Fiore}, {Menci}, {Cresci}, {La Franca}, \&
		{Merloni}}]{BongiornoMaiolinoBrusaEtAl2014}
	{Bongiorno}, A., {Maiolino}, R., {Brusa}, M., {et~al.} 2014, \mnras, 443, 2077
	
	\bibitem[{{Boroson} \& {Green}(1992)}]{BorosonGreen1992}
	{Boroson}, T.~A. \& {Green}, R.~F. 1992, \apjs, 80, 109
	
	\bibitem[{{Bouch{\'e}} {et~al.}(2007){Bouch{\'e}}, {Cresci}, {Davies},
		{Eisenhauer}, {F{\"o}rster Schreiber}, {Genzel}, {Gillessen}, {Lehnert},
		{Lutz}, {Nesvadba}, {Shapiro}, {Sternberg}, {Tacconi}, {Verma}, {Cimatti},
		{Daddi}, {Renzini}, {Erb}, {Shapley}, \&
		{Steidel}}]{BoucheCresciDaviesEtAl2007}
	{Bouch{\'e}}, N., {Cresci}, G., {Davies}, R., {et~al.} 2007, \apj, 671, 303
	
	\bibitem[{{Brusa} {et~al.}(2015){Brusa}, {Bongiorno}, {Cresci}, {Perna},
		{Marconi}, {Mainieri}, {Maiolino}, {Salvato}, {Lusso}, {Santini}, {Comastri},
		{Fiore}, {Gilli}, {La Franca}, {Lanzuisi}, {Lutz}, {Merloni}, {Mignoli},
		{Onori}, {Piconcelli}, {Rosario}, {Vignali}, \&
		{Zamorani}}]{BrusaBongiornoCresciEtAl2015}
	{Brusa}, M., {Bongiorno}, A., {Cresci}, G., {et~al.} 2015, \mnras, 446, 2394
	
	\bibitem[{{Brusa} {et~al.}(2016){Brusa}, {Perna}, {Cresci}, {Schramm},
		{Delvecchio}, {Lanzuisi}, {Mainieri}, {Mignoli}, {Zamorani}, {Berta},
		{Bongiorno}, {Comastri}, {Fiore}, {Kakkad}, {Marconi}, {Rosario}, {Contini},
		\& {Lamareille}}]{BrusaPernaCresciEtAl2016}
	{Brusa}, M., {Perna}, M., {Cresci}, G., {et~al.} 2016, \aap, 588, A58
	
	\bibitem[{{Bruzual} \& {Charlot}(2003)}]{BruzualCharlot2003}
	{Bruzual}, G. \& {Charlot}, S. 2003, \mnras, 344, 1000
	
	\bibitem[{{Calzetti} {et~al.}(2000){Calzetti}, {Armus}, {Bohlin}, {Kinney},
		{Koornneef}, \& {Storchi-Bergmann}}]{CalzettiArmusBohlinEtAl2000}
	{Calzetti}, D., {Armus}, L., {Bohlin}, R.~C., {et~al.} 2000, \apj, 533, 682
	
	\bibitem[{{Cano-D{\'{\i}}az} {et~al.}(2012){Cano-D{\'{\i}}az}, {Maiolino},
		{Marconi}, {Netzer}, {Shemmer}, \&
		{Cresci}}]{Cano-DiazMaiolinoMarconiEtAl2012}
	{Cano-D{\'{\i}}az}, M., {Maiolino}, R., {Marconi}, A., {et~al.} 2012, \aap,
	537, L8
	
	\bibitem[{{Carniani} {et~al.}(2015){Carniani}, {Marconi}, {Maiolino},
		{Balmaverde}, {Brusa}, {Cano-D{\'{\i}}az}, {Cicone}, {Comastri}, {Cresci},
		{Fiore}, {Feruglio}, {La Franca}, {Mainieri}, {Mannucci}, {Nagao}, {Netzer},
		{Piconcelli}, {Risaliti}, {Schneider}, \&
		{Shemmer}}]{CarnianiMarconiMaiolinoEtAl2015}
	{Carniani}, S., {Marconi}, A., {Maiolino}, R., {et~al.} 2015, \aap, 580, A102
	
	\bibitem[{{Carniani} {et~al.}(2016){Carniani}, {Marconi}, {Maiolino},
		{Balmaverde}, {Brusa}, {Cano-D{\'{\i}}az}, {Cicone}, {Comastri}, {Cresci},
		{Fiore}, {Feruglio}, {La Franca}, {Mainieri}, {Mannucci}, {Nagao}, {Netzer},
		{Piconcelli}, {Risaliti}, {Schneider}, \&
		{Shemmer}}]{CarnianiMarconiMaiolinoEtAl2016}
	{Carniani}, S., {Marconi}, A., {Maiolino}, R., {et~al.} 2016, \aap, 591, A28
	
	\bibitem[{{Cicone} {et~al.}(2014){Cicone}, {Maiolino}, {Sturm},
		{Graci{\'a}-Carpio}, {Feruglio}, {Neri}, {Aalto}, {Davies}, {Fiore},
		{Fischer}, {Garc{\'{\i}}a-Burillo}, {Gonz{\'a}lez-Alfonso},
		{Hailey-Dunsheath}, {Piconcelli}, \&
		{Veilleux}}]{CiconeMaiolinoSturmEtAl2014}
	{Cicone}, C., {Maiolino}, R., {Sturm}, E., {et~al.} 2014, \aap, 562, A21
	
	\bibitem[{{Costa} {et~al.}(2014){Costa}, {Sijacki}, \&
		{Haehnelt}}]{CostaSijackiHaehnelt2014}
	{Costa}, T., {Sijacki}, D., \& {Haehnelt}, M.~G. 2014, \mnras, 444, 2355
	
	\bibitem[{{Cresci} {et~al.}(2015){Cresci}, {Mainieri}, {Brusa}, {Marconi},
		{Perna}, {Mannucci}, {Piconcelli}, {Maiolino}, {Feruglio}, {Fiore},
		{Bongiorno}, {Lanzuisi}, {Merloni}, {Schramm}, {Silverman}, \&
		{Civano}}]{CresciMainieriBrusaEtAl2015}
	{Cresci}, G., {Mainieri}, V., {Brusa}, M., {et~al.} 2015, \apj, 799, 82
	
	\bibitem[{{Di Matteo} {et~al.}(2005){Di Matteo}, {Springel}, \&
		{Hernquist}}]{DiMatteoSpringelHernquist2005}
	{Di Matteo}, T., {Springel}, V., \& {Hernquist}, L. 2005, \nat, 433, 604
	
	\bibitem[{{Fabian}(2012)}]{Fabian2012}
	{Fabian}, A.~C. 2012, \araa, 50, 455
	
	\bibitem[{{Faucher-Gigu{\`e}re} \&
		{Quataert}(2012)}]{FaucherGiguereQuataert2012}
	{Faucher-Gigu{\`e}re}, C.-A. \& {Quataert}, E. 2012, \mnras, 425, 605
	
	\bibitem[{{Ferland} {et~al.}(2013){Ferland}, {Porter}, {van Hoof}, {Williams},
		{Abel}, {Lykins}, {Shaw}, {Henney}, \& {Stancil}}]{FerlandPorterVanHoof2013}
	{Ferland}, G.~J., {Porter}, R.~L., {van Hoof}, P.~A.~M., {et~al.} 2013, \rmxaa,
	49, 137
	
	\bibitem[{{Feruglio} {et~al.}(2015){Feruglio}, {Fiore}, {Carniani},
		{Piconcelli}, {Zappacosta}, {Bongiorno}, {Cicone}, {Maiolino}, {Marconi},
		{Menci}, {Puccetti}, \& {Veilleux}}]{FeruglioFioreCarnianiEtAl2015}
	{Feruglio}, C., {Fiore}, F., {Carniani}, S., {et~al.} 2015, \aap, 583, A99
	
	\bibitem[{{Feruglio} {et~al.}(2010){Feruglio}, {Maiolino}, {Piconcelli},
		{Menci}, {Aussel}, {Lamastra}, \&
		{Fiore}}]{FeruglioMaiolinoPiconcelliEtAl2010}
	{Feruglio}, C., {Maiolino}, R., {Piconcelli}, E., {et~al.} 2010, \aap, 518,
	L155
	
	\bibitem[{{Fiore} {et~al.}(2016, A\&A submitted){Fiore}, {Feruglio}, {Shankar},
		{Bischetti}, {Bongiorno}, {Menci}, {Maiolino}, {Piconcelli}, {Vietri}, \&
		{Zappacosta}}]{FioreEtAl2016}
	{Fiore}, F., {Feruglio}, C., {Shankar}, F., {et~al.} 2016, A\&A submitted
	
	\bibitem[{{Gallerani} {et~al.}(2010){Gallerani}, {Maiolino}, {Juarez}, {Nagao},
		{Marconi}, {Bianchi}, {Schneider}, {Mannucci}, {Oliva}, {Willott}, {Jiang},
		\& {Fan}}]{GalleraniMaiolinoJuarezEtAl2010}
	{Gallerani}, S., {Maiolino}, R., {Juarez}, Y., {et~al.} 2010, \aap, 523, A85
	
	\bibitem[{{Genzel} {et~al.}(2014){Genzel}, {F{\"o}rster Schreiber}, {Rosario},
		{Lang}, {Lutz}, {Wisnioski}, {Wuyts}, {Wuyts}, {Bandara}, {Bender}, {Berta},
		{Kurk}, {Mendel}, {Tacconi}, {Wilman}, {Beifiori}, {Brammer}, {Burkert},
		{Buschkamp}, {Chan}, {Carollo}, {Davies}, {Eisenhauer}, {Fabricius},
		{Fossati}, {Kriek}, {Kulkarni}, {Lilly}, {Mancini}, {Momcheva}, {Naab},
		{Nelson}, {Renzini}, {Saglia}, {Sharples}, {Sternberg}, {Tacchella}, \& {van
			Dokkum}}]{GenzelFoersterSchreiberRosarioEtAl2014}
	{Genzel}, R., {F{\"o}rster Schreiber}, N.~M., {Rosario}, D., {et~al.} 2014,
	\apj, 796, 7
	
	\bibitem[{{Genzel} {et~al.}(2011){Genzel}, {Newman}, {Jones}, {F{\"o}rster
			Schreiber}, {Shapiro}, {Genel}, {Lilly}, {Renzini}, {Tacconi}, {Bouch{\'e}},
		{Burkert}, {Cresci}, {Buschkamp}, {Carollo}, {Ceverino}, {Davies}, {Dekel},
		{Eisenhauer}, {Hicks}, {Kurk}, {Lutz}, {Mancini}, {Naab}, {Peng},
		{Sternberg}, {Vergani}, \& {Zamorani}}]{GenzelNewmanJonesEtAl2011}
	{Genzel}, R., {Newman}, S., {Jones}, T., {et~al.} 2011, \apj, 733, 101
	
	\bibitem[{{Gibson} {et~al.}(2009){Gibson}, {Jiang}, {Brandt}, {Hall}, {Shen},
		{Wu}, {Anderson}, {Schneider}, {Vanden Berk}, {Gallagher}, {Fan}, \&
		{York}}]{GibsonJiangBrandtEtAl2009}
	{Gibson}, R.~R., {Jiang}, L., {Brandt}, W.~N., {et~al.} 2009, \apj, 692, 758
	
	\bibitem[{{Greene} {et~al.}(2009){Greene}, {Zakamska}, {Liu}, {Barth}, \&
		{Ho}}]{GreeneZakamskaLiuEtAl2009}
	{Greene}, J.~E., {Zakamska}, N.~L., {Liu}, X., {Barth}, A.~J., \& {Ho}, L.~C.
	2009, \apj, 702, 441
	
	\bibitem[{{Harrison} {et~al.}(2016){Harrison}, {Alexander}, {Mullaney},
		{Stott}, {Swinbank}, {Arumugam}, {Bauer}, {Bower}, {Bunker}, \&
		{Sharples}}]{HarrisonAlexanderMullaneyEtAl2016}
	{Harrison}, C.~M., {Alexander}, D.~M., {Mullaney}, J.~R., {et~al.} 2016,
	\mnras, 456, 1195
	
	\bibitem[{{Harrison} {et~al.}(2014){Harrison}, {Alexander}, {Mullaney}, \&
		{Swinbank}}]{HarrisonAlexanderMullaneyEtAl2014}
	{Harrison}, C.~M., {Alexander}, D.~M., {Mullaney}, J.~R., \& {Swinbank}, A.~M.
	2014, \mnras, 441, 3306
	
	\bibitem[{{Harrison} {et~al.}(2012){Harrison}, {Alexander}, {Swinbank},
		{Smail}, {Alaghband-Zadeh}, {Bauer}, {Chapman}, {Del Moro}, {Hickox},
		{Ivison}, {Men{\'e}ndez-Delmestre}, {Mullaney}, \&
		{Nesvadba}}]{HarrisonAlexanderSwinbankEtAl2012}
	{Harrison}, C.~M., {Alexander}, D.~M., {Swinbank}, A.~M., {et~al.} 2012,
	\mnras, 426, 1073
	
	\bibitem[{{Hopkins} {et~al.}(2008){Hopkins}, {Hernquist}, {Cox}, \& {Kere{\v
				s}}}]{HopkinsHernquistCoxEtAl2008}
	{Hopkins}, P.~F., {Hernquist}, L., {Cox}, T.~J., \& {Kere{\v s}}, D. 2008,
	\apjs, 175, 356
	
	\bibitem[{{Hopkins} {et~al.}(2004){Hopkins}, {Strauss}, {Hall}, {Richards},
		{Cooper}, {Schneider}, {Vanden Berk}, {Jester}, {Brinkmann}, \&
		{Szokoly}}]{HopkinsStraussHallEtAl2004}
	{Hopkins}, P.~F., {Strauss}, M.~A., {Hall}, P.~B., {et~al.} 2004, \aj, 128,
	1112
	
	\bibitem[{{Hopkins} {et~al.}(2016){Hopkins}, {Torrey}, {Faucher-Gigu{\`e}re},
		{Quataert}, \& {Murray}}]{HopkinsTorreyFaucherGiguereEtAl2016}
	{Hopkins}, P.~F., {Torrey}, P., {Faucher-Gigu{\`e}re}, C.-A., {Quataert}, E.,
	\& {Murray}, N. 2016, \mnras, 458, 816
	
	\bibitem[{{Kakkad} {et~al.}(2016){Kakkad}, {Mainieri}, {Padovani}, {Cresci},
		{Husemann}, {Carniani}, {Brusa}, {Lamastra}, {Lanzuisi}, {Piconcelli}, \&
		{Schramm}}]{KakkadMainieriPadovaniEtAl2016}
	{Kakkad}, D., {Mainieri}, V., {Padovani}, P., {et~al.} 2016, \aap, 592, A148
	
	\bibitem[{{Kelly} \& {Merloni}(2012)}]{KellyMerloni2012}
	{Kelly}, B.~C. \& {Merloni}, A. 2012, Advances in Astronomy, 2012, 970858
	
	\bibitem[{{Kim} {et~al.}(2013){Kim}, {Ho}, {Lonsdale}, {Lacy}, {Blain}, \&
		{Kimball}}]{KimHoLonsdaleEtAl2013}
	{Kim}, M., {Ho}, L.~C., {Lonsdale}, C.~J., {et~al.} 2013, \apjl, 768, L9
	
	\bibitem[{{King} \& {Pounds}(2015)}]{KingPounds2015}
	{King}, A. \& {Pounds}, K. 2015, \araa, 53, 115
	
	\bibitem[{{Liu} {et~al.}(2013){Liu}, {Zakamska}, {Greene}, {Nesvadba}, \&
		{Liu}}]{LiuZakamskaGreeneEtAl2013}
	{Liu}, G., {Zakamska}, N.~L., {Greene}, J.~E., {Nesvadba}, N.~P.~H., \& {Liu},
	X. 2013, \mnras, 436, 2576
	
	\bibitem[{{Lusso} {et~al.}(2012){Lusso}, {Comastri}, {Simmons}, {Mignoli},
		{Zamorani}, {Vignali}, {Brusa}, {Shankar}, {Lutz}, {Trump}, {Maiolino},
		{Gilli}, {Bolzonella}, {Puccetti}, {Salvato}, {Impey}, {Civano}, {Elvis},
		{Mainieri}, {Silverman}, {Koekemoer}, {Bongiorno}, {Merloni}, {Berta}, {Le
			Floc'h}, {Magnelli}, {Pozzi}, \& {Riguccini}}]{LussoComastriSimmonsEtAl2012}
	{Lusso}, E., {Comastri}, A., {Simmons}, B.~D., {et~al.} 2012, \mnras, 425, 623
	
	\bibitem[{{Maiolino} {et~al.}(2012){Maiolino}, {Gallerani}, {Neri}, {Cicone},
		{Ferrara}, {Genzel}, {Lutz}, {Sturm}, {Tacconi}, {Walter}, {Feruglio},
		{Fiore}, \& {Piconcelli}}]{MaiolinoGalleraniNeriEtAl2012}
	{Maiolino}, R., {Gallerani}, S., {Neri}, R., {et~al.} 2012, \mnras, 425, L66
	
	\bibitem[{{Markwardt}(2009)}]{Markwardt2009}
	{Markwardt}, C.~B. 2009, in Astronomical Society of the Pacific Conference
	Series, Vol. 411, Astronomical Data Analysis Software and Systems XVIII, ed.
	D.~A. {Bohlender}, D.~{Durand}, \& P.~{Dowler}, 251
	
	\bibitem[{{Menci} {et~al.}(2008){Menci}, {Fiore}, {Puccetti}, \&
		{Cavaliere}}]{MenciFiorePuccettiEtAl2008}
	{Menci}, N., {Fiore}, F., {Puccetti}, S., \& {Cavaliere}, A. 2008, \apj, 686,
	219
	
	\bibitem[{{Morganti} {et~al.}(2016){Morganti}, {Veilleux}, {Oosterloo}, {Teng},
		\& {Rupke}}]{MorgantiVeilleuxOosterlooEtAl2016}
	{Morganti}, R., {Veilleux}, S., {Oosterloo}, T., {Teng}, S.~H., \& {Rupke}, D.
	2016, ArXiv e-prints [\eprint[arXiv]{1606.01640}]
	
	\bibitem[{{Mullaney} {et~al.}(2013){Mullaney}, {Alexander}, {Fine}, {Goulding},
		{Harrison}, \& {Hickox}}]{MullaneyAlexanderFineEtAl2013}
	{Mullaney}, J.~R., {Alexander}, D.~M., {Fine}, S., {et~al.} 2013, \mnras, 433,
	622
	
	\bibitem[{{Nagao} {et~al.}(2006){Nagao}, {Marconi}, \&
		{Maiolino}}]{NagaoMarconiMaiolino2006}
	{Nagao}, T., {Marconi}, A., \& {Maiolino}, R. 2006, \aap, 447, 157
	
	\bibitem[{{Nesvadba} {et~al.}(2008){Nesvadba}, {Lehnert}, {De Breuck},
		{Gilbert}, \& {van Breugel}}]{NesvadbaLehnertDeBreuckEtAl2008}
	{Nesvadba}, N.~P.~H., {Lehnert}, M.~D., {De Breuck}, C., {Gilbert}, A.~M., \&
	{van Breugel}, W. 2008, \aap, 491, 407
	
	\bibitem[{{Nesvadba} {et~al.}(2006){Nesvadba}, {Lehnert}, {Eisenhauer},
		{Gilbert}, {Tecza}, \& {Abuter}}]{NesvadbaLehnertEisenhauerEtAl2006}
	{Nesvadba}, N.~P.~H., {Lehnert}, M.~D., {Eisenhauer}, F., {et~al.} 2006, \apj,
	650, 693
	
	\bibitem[{{Perna} {et~al.}(2015{\natexlab{a}}){Perna}, {Brusa}, {Cresci},
		{Comastri}, {Lanzuisi}, {Lusso}, {Marconi}, {Salvato}, {Zamorani},
		{Bongiorno}, {Mainieri}, {Maiolino}, \& {Mignoli}}]{PernaBrusaCresciEtAl2015}
	{Perna}, M., {Brusa}, M., {Cresci}, G., {et~al.} 2015{\natexlab{a}}, \aap, 574,
	A82
	
	\bibitem[{{Perna} {et~al.}(2015{\natexlab{b}}){Perna}, {Brusa}, {Salvato},
		{Cresci}, {Lanzuisi}, {Berta}, {Delvecchio}, {Fiore}, {Lutz}, {Le Floc'h},
		{Mainieri}, \& {Riguccini}}]{PernaBrusaSalvatoEtAl2015}
	{Perna}, M., {Brusa}, M., {Salvato}, M., {et~al.} 2015{\natexlab{b}}, \aap,
	583, A72
	
	\bibitem[{{Peterson}(1997)}]{Peterson1997}
	{Peterson}, B.~M. 1997, {An Introduction to Active Galactic Nuclei}
	
	\bibitem[{{Prevot} {et~al.}(1984){Prevot}, {Lequeux}, {Prevot}, {Maurice}, \&
		{Rocca-Volmerange}}]{PrevotLequeuxPrevotEtAl1984}
	{Prevot}, M.~L., {Lequeux}, J., {Prevot}, L., {Maurice}, E., \&
	{Rocca-Volmerange}, B. 1984, \aap, 132, 389
	
	\bibitem[{{Richards} {et~al.}(2003){Richards}, {Hall}, {Vanden Berk},
		{Strauss}, {Schneider}, {Weinstein}, {Reichard}, {York}, {Knapp}, {Fan},
		{Ivezi{\'c}}, {Brinkmann}, {Budav{\'a}ri}, {Csabai}, \&
		{Nichol}}]{RichardsHallVandenBerkEtAl2003}
	{Richards}, G.~T., {Hall}, P.~B., {Vanden Berk}, D.~E., {et~al.} 2003, \aj,
	126, 1131
	
	\bibitem[{{Richards} {et~al.}(2006){Richards}, {Lacy}, {Storrie-Lombardi},
		{Hall}, {Gallagher}, {Hines}, {Fan}, {Papovich}, {Vanden Berk}, {Trammell},
		{Schneider}, {Vestergaard}, {York}, {Jester}, {Anderson}, {Budav{\'a}ri}, \&
		{Szalay}}]{RichardsLacyStorrie-LombardiEtAl2006}
	{Richards}, G.~T., {Lacy}, M., {Storrie-Lombardi}, L.~J., {et~al.} 2006, \apjs,
	166, 470
	
	\bibitem[{{Rieke} \& {Lebofsky}(1985)}]{RiekeLebofsky1985}
	{Rieke}, G.~H. \& {Lebofsky}, M.~J. 1985, \apj, 288, 618
	
	\bibitem[{{Rodr{\'{\i}}guez Zaur{\'{\i}}n} {et~al.}(2013){Rodr{\'{\i}}guez
			Zaur{\'{\i}}n}, {Tadhunter}, {Rose}, \&
		{Holt}}]{RodriguezZaurinTadhunterEtAl2013}
	{Rodr{\'{\i}}guez Zaur{\'{\i}}n}, J., {Tadhunter}, C.~N., {Rose}, M., \&
	{Holt}, J. 2013, \mnras, 432, 138
	
	\bibitem[{{Rupke} {et~al.}(2002){Rupke}, {Veilleux}, \&
		{Sanders}}]{RupkeVeilleux2002}
	{Rupke}, D.~S., {Veilleux}, S., \& {Sanders}, D.~B. 2002, \apj, 570, 588
	
	\bibitem[{{Rupke} {et~al.}(2005){Rupke}, {Veilleux}, \&
		{Sanders}}]{RupkeVeilleuxSanders2005}
	{Rupke}, D.~S., {Veilleux}, S., \& {Sanders}, D.~B. 2005, \apjs, 160, 115
	
	\bibitem[{{Rupke} \& {Veilleux}(2013)}]{RupkeVeilleux2013}
	{Rupke}, D.~S.~N. \& {Veilleux}, S. 2013, \apj, 768, 75
	
	\bibitem[{{Rupke} \& {Veilleux}(2015)}]{RupkeVeilleux2015}
	{Rupke}, D.~S.~N. \& {Veilleux}, S. 2015, \apj, 801, 126
	
	\bibitem[{{Shemmer} {et~al.}(2004){Shemmer}, {Netzer}, {Maiolino}, {Oliva},
		{Croom}, {Corbett}, \& {di Fabrizio}}]{ShemmerNetzerMaiolinoEtAl2004}
	{Shemmer}, O., {Netzer}, H., {Maiolino}, R., {et~al.} 2004, \apj, 614, 547
	
	\bibitem[{{Shen}(2016)}]{Shen2016}
	{Shen}, Y. 2016, \apj, 817, 55
	
	\bibitem[{{Shen} \& {Ho}(2014)}]{ShenHo2014}
	{Shen}, Y. \& {Ho}, L.~C. 2014, \nat, 513, 210
	
	\bibitem[{{Shen} \& {Liu}(2012)}]{ShenLiu2012}
	{Shen}, Y. \& {Liu}, X. 2012, \apj, 753, 125
	
	\bibitem[{{Silk} \& {Rees}(1998)}]{SilkRees1998}
	{Silk}, J. \& {Rees}, M.~J. 1998, \aap, 331, L1
	
	\bibitem[{{Skrutskie} {et~al.}(2006){Skrutskie}, {Cutri}, {Stiening},
		{Weinberg}, {Schneider}, {Carpenter}, {Beichman}, {Capps}, {Chester},
		{Elias}, {Huchra}, {Liebert}, {Lonsdale}, {Monet}, {Price}, {Seitzer},
		{Jarrett}, {Kirkpatrick}, {Gizis}, {Howard}, {Evans}, {Fowler}, {Fullmer},
		{Hurt}, {Light}, {Kopan}, {Marsh}, {McCallon}, {Tam}, {Van Dyk}, \&
		{Wheelock}}]{StruskieCutriStieningEtAl2006}
	{Skrutskie}, M.~F., {Cutri}, R.~M., {Stiening}, R., {et~al.} 2006, \aj, 131,
	1163
	
	\bibitem[{{Somerville} {et~al.}(2008){Somerville}, {Hopkins}, {Cox},
		{Robertson}, \& {Hernquist}}]{SomervilleHopkinsCoxEtAl2008}
	{Somerville}, R.~S., {Hopkins}, P.~F., {Cox}, T.~J., {Robertson}, B.~E., \&
	{Hernquist}, L. 2008, \mnras, 391, 481
	
	\bibitem[{{Spoon} {et~al.}(2013){Spoon}, {Farrah}, {Lebouteiller},
		{Gonz{\'a}lez-Alfonso}, {Bernard-Salas}, {Urrutia}, {Rigopoulou},
		{Westmoquette}, {Smith}, {Afonso}, {Pearson}, {Cormier}, {Efstathiou},
		{Borys}, {Verma}, {Etxaluze}, \&
		{Clements}}]{SpoonFarrahLebouteillerEtAl2013}
	{Spoon}, H.~W.~W., {Farrah}, D., {Lebouteiller}, V., {et~al.} 2013, \apj, 775,
	127
	
	\bibitem[{{Tsuzuki} {et~al.}(2006){Tsuzuki}, {Kawara}, {Yoshii}, {Oyabu},
		{Tanab{\'e}}, \& {Matsuoka}}]{TsuzukiKawaraYoshiiEtAl2006}
	{Tsuzuki}, Y., {Kawara}, K., {Yoshii}, Y., {et~al.} 2006, \apj, 650, 57
	
	\bibitem[{{Urrutia} {et~al.}(2012){Urrutia}, {Lacy}, {Spoon}, {Glikman},
		{Petric}, \& {Schulz}}]{UrrutiaLacySpoonEtAl2012}
	{Urrutia}, T., {Lacy}, M., {Spoon}, H., {et~al.} 2012, \apj, 757, 125
	
	\bibitem[{{Vacca} {et~al.}(2003){Vacca}, {Cushing}, \&
		{Rayner}}]{VaccaCushingRayner2003}
	{Vacca}, W.~D., {Cushing}, M.~C., \& {Rayner}, J.~T. 2003, \pasp, 115, 389
	
	\bibitem[{{Vanden Berk} {et~al.}(2001){Vanden Berk}, {Richards}, {Bauer},
		{Strauss}, {Schneider}, {Heckman}, {York}, {Hall}, {Fan}, {Knapp},
		{Anderson}, {Annis}, {Bahcall}, {Bernardi}, {Briggs}, {Brinkmann}, {Brunner},
		{Burles}, {Carey}, {Castander}, {Connolly}, {Crocker}, {Csabai}, {Doi},
		{Finkbeiner}, {Friedman}, {Frieman}, {Fukugita}, {Gunn}, {Hennessy},
		{Ivezi{\'c}}, {Kent}, {Kunszt}, {Lamb}, {Leger}, {Long}, {Loveday}, {Lupton},
		{Meiksin}, {Merelli}, {Munn}, {Newberg}, {Newcomb}, {Nichol}, {Owen}, {Pier},
		{Pope}, {Rockosi}, {Schlegel}, {Siegmund}, {Smee}, {Snir}, {Stoughton},
		{Stubbs}, {SubbaRao}, {Szalay}, {Szokoly}, {Tremonti}, {Uomoto}, {Waddell},
		{Yanny}, \& {Zheng}}]{VandenBerkRichardsBauer2001}
	{Vanden Berk}, D.~E., {Richards}, G.~T., {Bauer}, A., {et~al.} 2001, \aj, 122,
	549
	
	\bibitem[{{V{\'e}ron-Cetty} {et~al.}(2004){V{\'e}ron-Cetty}, {Joly}, \&
		{V{\'e}ron}}]{Veron-CettyJollyVeron2004}
	{V{\'e}ron-Cetty}, M.-P., {Joly}, M., \& {V{\'e}ron}, P. 2004, \aap, 417, 515
	
	\bibitem[{{Villar-Mart{\'{\i}}n} {et~al.}(2011){Villar-Mart{\'{\i}}n},
		{Humphrey}, {Delgado}, {Colina}, \&
		{Arribas}}]{Villar-MartinHumphreyDelgadoEtAl2011}
	{Villar-Mart{\'{\i}}n}, M., {Humphrey}, A., {Delgado}, R.~G., {Colina}, L., \&
	{Arribas}, S. 2011, \mnras, 418, 2032
	
	\bibitem[{{Weedman} {et~al.}(2012){Weedman}, {Sargsyan}, {Lebouteiller},
		{Houck}, \& {Barry}}]{WeedmanSargsyanLebouteillerEtAl2012}
	{Weedman}, D., {Sargsyan}, L., {Lebouteiller}, V., {Houck}, J., \& {Barry}, D.
	2012, \apj, 761, 184
	
	\bibitem[{{Wright} {et~al.}(2010){Wright}, {Eisenhardt}, {Mainzer}, {Ressler},
		{Cutri}, {Jarrett}, {Kirkpatrick}, {Padgett}, {McMillan}, {Skrutskie},
		{Stanford}, {Cohen}, {Walker}, {Mather}, {Leisawitz}, {Gautier}, {McLean},
		{Benford}, {Lonsdale}, {Blain}, {Mendez}, {Irace}, {Duval}, {Liu}, {Royer},
		{Heinrichsen}, {Howard}, {Shannon}, {Kendall}, {Walsh}, {Larsen}, {Cardon},
		{Schick}, {Schwalm}, {Abid}, {Fabinsky}, {Naes}, \&
		{Tsai}}]{WrightEisenhardtMainzerEtAl2010}
	{Wright}, E.~L., {Eisenhardt}, P.~R.~M., {Mainzer}, A.~K., {et~al.} 2010, \aj,
	140, 1868
	
	\bibitem[{{Wyithe} \& {Loeb}(2003)}]{WyitheLoeb2003}
	{Wyithe}, J.~S.~B. \& {Loeb}, A. 2003, \apj, 595, 614
	
	\bibitem[{{Wylezalek} \& {Zakamska}(2016)}]{WylezalekZakamska2016}
	{Wylezalek}, D. \& {Zakamska}, N.~L. 2016, \mnras, 461, 3724
	
	\bibitem[{{Zakamska} \& {Greene}(2014)}]{ZakamskaGreene2014}
	{Zakamska}, N.~L. \& {Greene}, J.~E. 2014, \mnras, 442, 784
	
	\bibitem[{{Zakamska} {et~al.}(2016){Zakamska}, {Hamann}, {P{\^a}ris}, {Brandt},
		{Greene}, {Strauss}, {Villforth}, {Wylezalek}, {Alexandroff}, \&
		{Ross}}]{ZakamskaHamannParisEtAl.2016}
	{Zakamska}, N.~L., {Hamann}, F., {P{\^a}ris}, I., {et~al.} 2016, \mnras, 459,
	3144
	
	\bibitem[{Zhang {et~al.}(2011)Zhang, Dong, Wang, \&
		Gaskell}]{ZhangDongWangEtAl2011}
	Zhang, K., Dong, X.-B., Wang, T.-G., \& Gaskell, C.~M. 2011, The Astrophysical
	Journal, 737, 71
	
	\bibitem[{{Zubovas} \& {King}(2012)}]{ZubovasKing2012}
	{Zubovas}, K. \& {King}, A. 2012, \apjl, 745, L34
	
\end{thebibliography}
\end{document}